%% file: main.tex
\documentclass[twocolumn,prl,superscriptaddress]{revtex4-1}

\usepackage[dvipsnames]{xcolor}
\usepackage{graphicx}
\usepackage{amssymb}
\usepackage{amsmath}
\usepackage{amsfonts}
\usepackage{mathtools}
\usepackage{tikz}
\usepackage[normalem]{ulem}

\usepackage{bm}

\usepackage[colorlinks=true,
            linkcolor=red,
            citecolor=blue,
            urlcolor=blue]{hyperref}

\newcommand{\sgn}{\text{sgn}}

\newcommand{\bea}{\begin{equation}\begin{aligned}}
\newcommand{\eea}{\end{aligned}\end{equation}}

\DeclareMathOperator\arctanh{arctanh}

\begin{document}

\newcommand{\mytitle}{Spread and erase -- How electron hydrodynamics can eliminate the Landauer-Sharvin resistance}

\title{\mytitle}

\author{Ady Stern}
\affiliation{Department of Condensed Matter Physics, Weizmann Institute of Science, Rehovot 76100, Israel}
\author{Thomas Scaffidi}
\affiliation{Department of Physics and Astronomy, University of California, Irvine, California 92697, USA}
\affiliation{Department of Physics, University of Toronto, 60 St. George Street, Toronto, Ontario, M5S 1A7, Canada}
\author{Oren Reuven}
\affiliation{Department of Condensed Matter Physics, Weizmann Institute of Science, Rehovot 76100, Israel}
\author{Chandan Kumar}
\affiliation{Department of Condensed Matter Physics, Weizmann Institute of Science, Rehovot 76100, Israel}
\author{John Birkbeck}
\affiliation{Department of Condensed Matter Physics, Weizmann Institute of Science, Rehovot 76100, Israel}
\author{Shahal Ilani}
\affiliation{Department of Condensed Matter Physics, Weizmann Institute of Science, Rehovot 76100, Israel}

\begin{abstract}
It has long been realized that even a perfectly clean electronic system harbors a Landauer-Sharvin resistance, inversely proportional to the number of its conduction channels. This resistance is usually associated with voltage drops on the system's contacts to an external circuit. Recent theories have shown that hydrodynamic effects can reduce this resistance, raising the question of the lower bound of resistance of hydrodynamic electrons. Here we show that by a proper choice of device geometry, it is possible to spread the Landauer-Sharvin resistance throughout the bulk of the system, allowing its complete elimination by electron hydrodynamics. We trace the effect to the dynamics of electrons flowing in channels that terminate within the sample. For ballistic systems this termination leads to back-reflection of the electrons and creates resistance. Hydrodynamically, the scattering of these electrons off other electrons allows them to transfer to transmitted channels and avoid the resistance. Counter-intuitively, we find that in contrast to the ohmic regime, for hydrodynamic electrons the resistance of a device with a given width can decrease with its length, suggesting that a long enough device may have an arbitrarily small total resistance.
\end{abstract}

\maketitle

{\it Introduction} The electronic resistivity to the flow of current is a fundamental quantity in condensed matter physics. Frequently, its minimization is desired. The Drude model, dating back to 1900, suggests that the resistivity originates mostly from   momentum loss to impurities. However, it was realized that even in the ballistic limit, in which impurities and phonons are absent, the interface between the electronic system and the metallic contacts to which it is coupled carries another fundamental source of resistance - the Landauer-Sharvin (LS) resistance\cite{Landauer1957,Sharvin1965,datta_1995,imry2002introduction}. This resistance is inversely proportional to the number of quantum mechanical channels that are transmitted through the system. 

More recently,  another regime of transport was discovered, in which electrons behave like a viscous fluid due to strong momentum-conserving electron-electron scattering. 
\cite{gurzhi1963minimum,gurzhi1968hydrodynamic,PhysRevB.21.3279,PhysRevLett.52.368,gurzhi1995electron,PhysRevB.49.5038,PhysRevB.51.13389,PhysRevLett.52.368,PhysRevLett.71.2465,PhysRevLett.77.1143,Spivak20062071,PhysRevLett.106.256804,PhysRevB.92.165433,PhysRevLett.117.166601,PhysRevLett.113.235901,levitov,PhysRevB.92.165433,PhysRevB.93.125410,PhysRevB.94.125427,2016arXiv161209239G,PhysRevB.95.115425,PhysRevB.95.121301,Bandurin1055,Crossno1058,Moll1061,Narozhny2017,Guo3068,Kumar2017,Scaffidi2017,PhysRevB.97.045105,PhysRevB.97.121404,PhysRevLett.121.176805,PhysRevB.98.165412,PhysRevB.97.121405,Gooth2018,Braem2018,Berdyugin2019,Sulpizio2019,NAROZHNY2019167979,PhysRevLett.123.026801,Ku2020,LEVCHENKO2020168218,PhysRevB.100.245305,Jenkins2020,Keser2021,Gupta2021,Krebs2021,Hong2020}.
Somewhat counter-intuitively, it was shown that the resistance in this hydrodynamical regime may be lower than the ballistic one, suggesting the term  ``super-ballistic''\cite{Nagaev2008,Nagaev2010,Melnikov2012,Guo3068,Kumar2017}. Furthermore, conditions in which field-free current flow may locally exist were suggested\cite{PhysRevLett.123.026801}.

\begin{figure}[t]
\includegraphics[width=0.52\textwidth,clip]{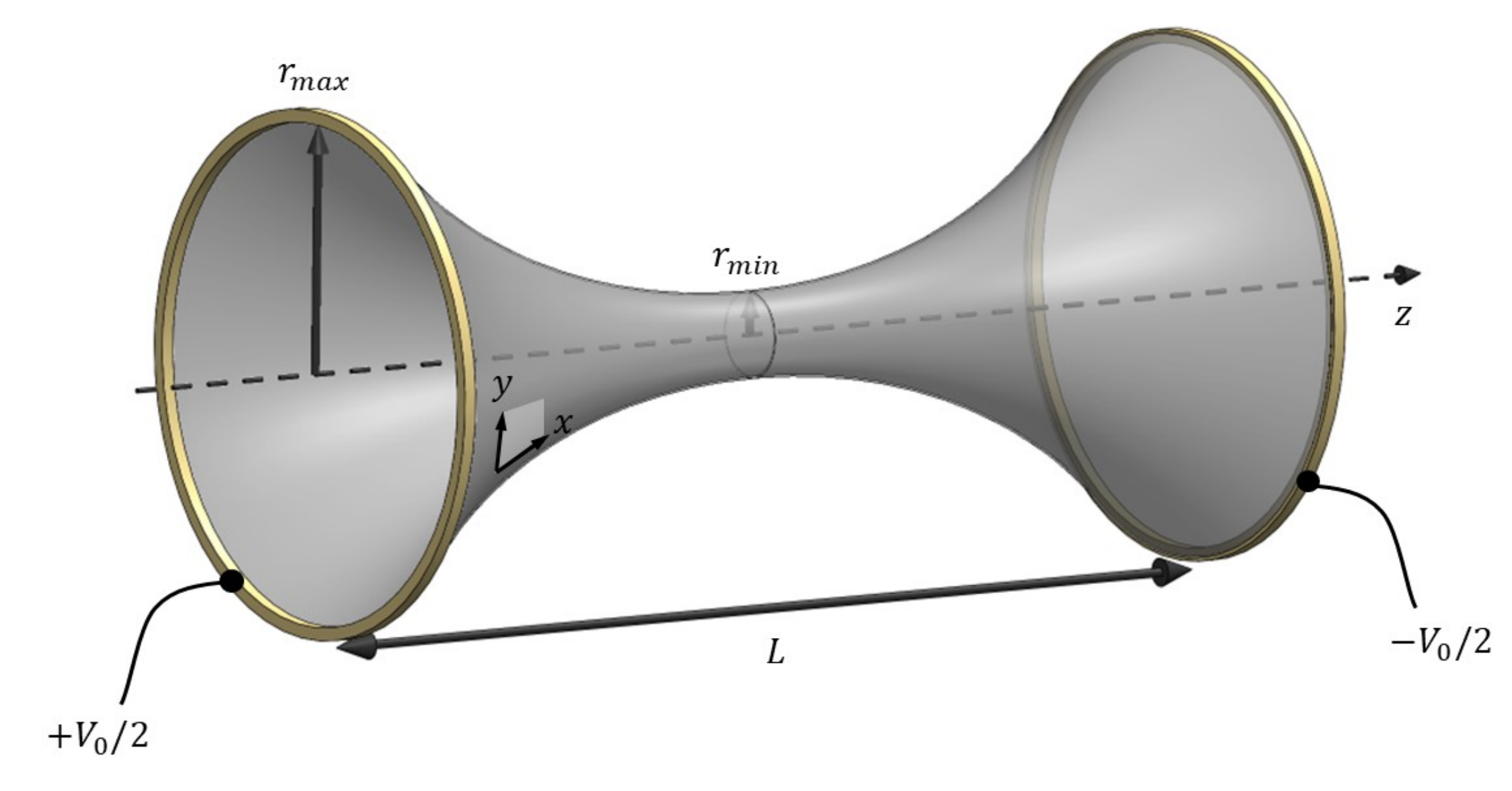} 
\caption{\label{wormhole-fig} The wormhole geometry is a two dimensional azimuthally-symmetric electronic system embedded in three dimensional space, described by the equation $r=r(z)$, where the radius is maximal ($r_{max}$) at the interface to the contacts and minimal ($r_{min}$) at the center. A current $I$ is driven from negative to positive $z$, and the potentials at the two contacts are $V(\mp L/2)=\pm \frac{V_0}{2}$.  }
\end{figure}

In this work, using a combination of Landauer and Boltzmann analyses we demonstrate a mechanism by which electron hydrodynamics can eliminate the LS resistance, and find the minimal value that this resistance may attain. Our study is semi-classical and focuses on two dimensional systems. We describe an electronic system in terms of its conduction channels, and show that when the number of  channels varies along the direction of the current flow, the Landauer-Sharvin resistance detaches from the contacts and spreads over the bulk of the electronic system. When the length scale of this spreading is larger than the electron-electron scattering mean free path, $\ell_{ee}$, the resistance is dramatically suppressed. 

Microscopically, this suppression results from the scattering of electrons whose channels are being terminated due to a narrowing of the system's cross-section or a decrease of its carrier density. In a ballistic system, these electrons are reflected back and do not contribute to the current, thereby generating LS resistance in the sample's bulk. In contrast, in the hydrodynamic regime electron-electron scattering transfers these electrons into transmitted channels, thus avoiding their reflection and the corresponding resistance.

Equipped with this analysis, we can ask the question of the minimal resistance of hydrodynamic electrons flowing through a constriction.
In the ballistic case, for a sample of length $L$ and a minimal cross section $2\pi r_{min}$, the LS resistance is given by $\frac{h}{2e^2k_Fr_{min}}$($k_F$ is the Fermi momentum, and we consider a single spin species).
In the hydrodynamic case, previous works\cite{Guo3068,Kumar2017,PhysRevLett.123.026801} reported a reduction of the LS resistance by a factor of $\ell_{ee}/r_{min}$ due to electron hydrodynamics. We find a further reduction of the resistance by an additional factor of $r_{min}/L$ if the constriction's width varies over a scale $L\gg r_{min}$. In contrast to the familiar ohmic regime, in which resistance increases with $L$, the resistance in the hydrodynamic regime decreases with $L$. This implies that a system with a given $r_{min}$ and a large enough $L$ may have an arbitrarily small total resistance. 

{\it Wormhole geometry}  In order to study the resistance of hydrodynamic electrons in a generic expanding geometry while avoiding boundary effects, we use a ``wormhole'' geometry (Fig. (\ref{wormhole-fig})). This geometry is a two-dimensional surface of revolution embedded in three dimensions, with azimuthal symmetry (toward the end of the paper we consider also a Corbino disk and a bar with varying electronic density). In cylindrical coordinates  the wormhole is defined by $r=r(z)$, with its minimum radius,  $r_{\rm min}$, occurring at $z=0$, and maximum radius, $r_{max}$, occurring at the contacts positioned at $z=\pm L/2$. For simplicity we assume $r(z)=r(-z)$. A current $I$ driven through the wormhole in the $z$-direction leads to a potential $V(z=\mp L/2)=\pm \frac{V_0}{2}$ at its contacts. On the manifold, we define a local Cartesian system of  coordinates tangent to the manifold, in which ${\hat y}$ is the unit vector in the azimuthal direction, and $\hat x=\frac{1}{\sqrt{1+r'2}}(r',0,1)$ is the unit vector in the direction along the manifold. For brevity, 
 we set $\hbar=e=1$. 

{\it Boltzmann description} Time-independent transport in a wormhole geometry may be described by a Boltzmann equation. 
In the absence of a driving force,  the magnitude of the electron's momentum is  constant, but its direction varies. Consequently, the equation reads:
\begin{equation}
\cos\theta \partial_z f  - \frac{r'}{r} \sin\theta \partial_\theta f = \sqrt{1+r'^2} I[f]
    \label{Boltzmann}
\end{equation}
where $f({\bf r},{\bf p})$ is the deviation of the number of electrons in a position ${\bf r}$ with momentum ${\bf p}$ from its equilibrium value, $\theta$ is the angle of the momentum with respect to the locally defined $x$-direction, $r'\equiv dr/dz$, and $I[f]$ is the scattering integral, elaborated below.
As explained in the Appendix, this equation is derived in two steps. First, we solve for the trajectories of free particles constrained to the manifold. Second, we equate the variation of $f$ along these trajectories with the scattering integral $I[f]$.

It is common to substitute the ansatz 
\begin{equation}
    f({\bf p},{\bf r})=\delta(\epsilon_F-\epsilon({\bf p}))h({\bf p},{\bf r})
    \label{ansatz}
\end{equation}
in (\ref{Boltzmann}), and integrate both sides over the magnitude of the momentum $\int\frac{ pdp}{4\pi^2}$, with $p=|{\bf p}|$. This integration  fixes $|{\bf p}|=p_F$ such that $h$ becomes a function of ${\bf r}$ and $\theta$, which describes the non-equilibrium angular shape of the Fermi surface. The integration replaces the $\delta$-function in (\ref{ansatz}) by a density of states at the Fermi energy and angle $\theta$, $\nu(E_F,\theta)=\nu_F/2\pi$ (here $\nu_F$ is the density of states at the Fermi energy). The Boltzmann equation becomes an equation for $\nu_F h(\theta, {\bf r})/2\pi$. The azimuthal symmetry reduces the dependence on $\bf r$ to a dependence on $z$ only.

{\it Landauer description} In the Landauer picture, the system is composed of $2j_{max}+1$ channels, enumerated by their angular momentum $j=-j_{max}..0..j_{max}$, with $j_{max}=k_Fr_{\rm max}$. The angular momentum $j=p_y(z)r(z)$, with $p_y$ the momentum in the azimuthal direction. Each channel is characterized by transmission and reflection probabilities $T_j,R_j$ satisfying $T_j+R_j=1$. We assume $r(z)$ to vary slowly on the scale of the Fermi wavelength, such that in the absence of interactions, channels with $|j|<k_Fr_{\rm min}$ are fully transmitted, and all other channels are fully reflected. The reflection takes place at the classical turning point $r(z)=|j|/k_F$. The current flowing through the wormhole is $I=\frac{k_Fr_{min}}{\pi }V_0$, leading to a dimensionless LS resistance $R_{ballistic}=\pi/k_F r_{min}$.

{\it ``Landauerizing'' Boltzmann} We reformulate the Boltzmann equation to elucidate its relation to the Landauer picture. To that end, we express the shape of the Fermi surface in terms of  different variables - the channel angular momentum $j$, the direction of motion, right (R) or left (L), and the position, $z$. Semi-classically the angular momentum is a real number, which is quantized to an integer in Landauer's quantum mechanical analysis. Here, we think about it semi-classically. 

Two steps are needed to transform the Boltzmann equation from an equation for $h(\theta,z)$ to an equation for the occupation in terms of $j, z$ and direction of motion, which we denote by $h^j_{R,L}(z)$. First, we change the variables in Eq. (\ref{Boltzmann}). Second, the integral $\int \frac{pdp}{4\pi^2}$ should be replaced by an integral over the $x$-component of the momentum, namely $p_F\int \frac{dp_x}{2\pi}$, where the limits are given by $p_x=0$ and $p_x=\pm \infty$, for R,L respectively. The $\delta$-function in (\ref{ansatz}) is then replaced by a density of states at the Fermi level at a fixed $j=p_yr(z)$, 
\begin{equation}
{\nu^j}=\frac{\nu_F }{\sqrt{1-\left (\frac{j}{k_Fr(z)}\right )^2}}\Theta\left ({k_Fr(z)-|j|}\right )  
\end{equation}
This density of states is inversely proportional to the $x$--component of the velocity, as familiar from Landauer's analysis. The details of the transformation are given in the Appendix, but the outcome is quite expected from the conservation of angular momentum: 
\begin{equation}
    \pm{ \nu_F}\partial_z h^j_{R,L}( z)= \sqrt{1+r'^2} {\tilde I}[{h^j_{R,L}( z)}]
    \label{Boltzmanng1}
\end{equation}
where the $\pm$ refers to right and left moving electrons, and $\tilde{I}$ is the scattering term expressed as a functional of $h_{R,L}^j(z)$, derived below.

The electronic density $\rho(z)$, current density $J_x(z)$ and potential $V(z)$ are, 
\begin{eqnarray}
    \rho({ z})&=&\int \frac{d{\bf p}}{4\pi^2}f({\bf p},{ z})=\frac{1}{2\pi k_Fr(z)}\int dj{\nu^j}(z)\left [{h_R^j}+{h_L^j}\right ]\nonumber \\
      J_x({ z})&=&\int \frac{d{\bf p}}{4\pi^2}\frac{p_x}{m}f({\bf p},z)=\frac{1}{4\pi^2 r(z)}\int dj \left[h^j_R-h^j_L\right]\nonumber \\
      V(z)&=&\rho(z)/\nu_F
    \label{currentdensity}
\end{eqnarray}
where the limits of integration are over all angular momenta for which $\nu^j\ne 0$, i.e. from $-j_{max}$ to $j_{max}=k_F r_{max}$.

\begin{figure}[t]
\centering
\includegraphics[width=.48\textwidth,trim={2.6cm 1.9cm 2cm 0 0},clip]{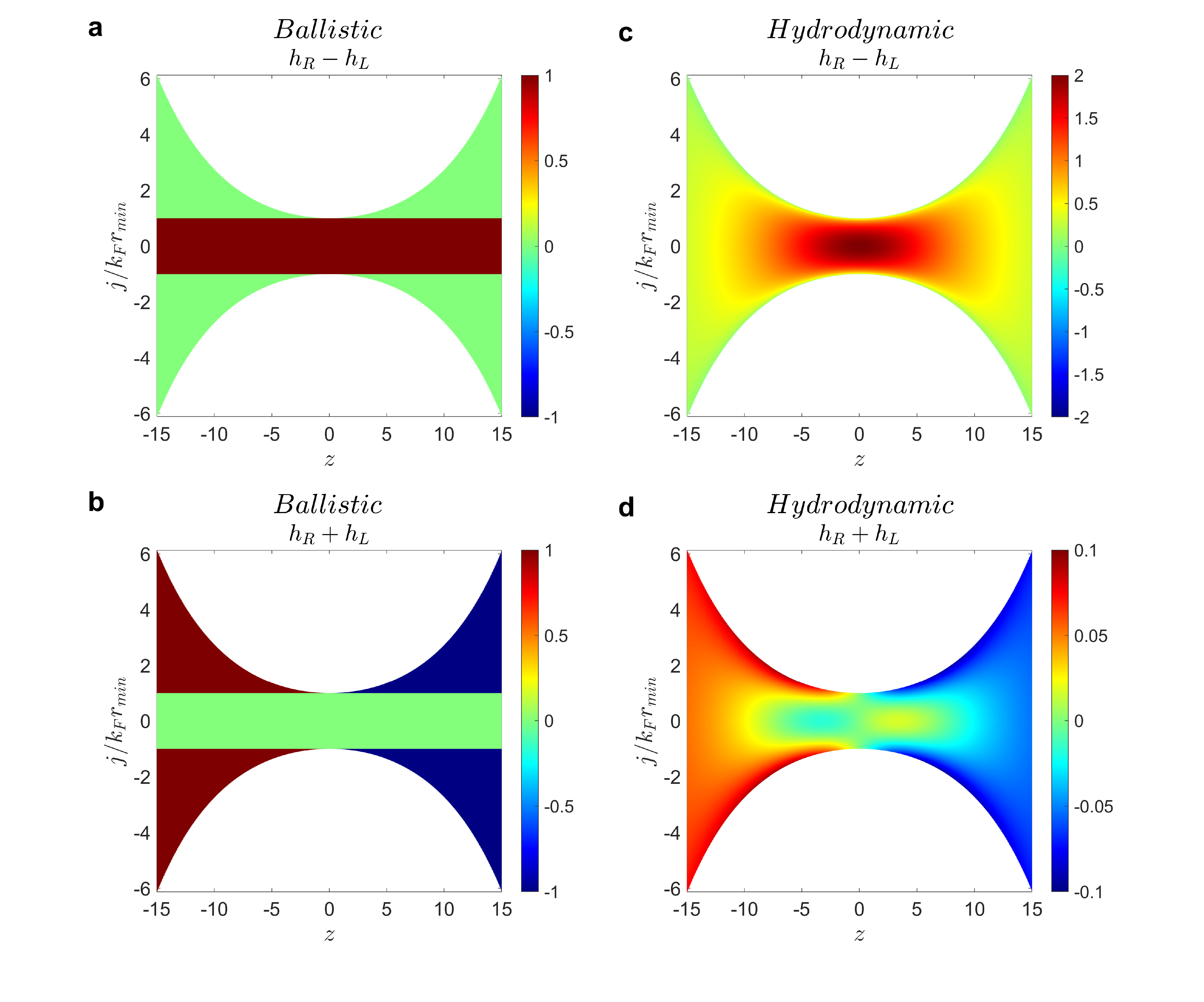}
\caption{\label{coshresults} Non-equilibrium distribution functions $h_R-h_L$ and $h_R+h_L$ for ballistic (a,b) and hydrodynamic (c,d) cases. These  distribution functions  contribute to the current density and voltage respectively (see Eq. (\ref{currentdensity})). They are plotted for the wormhole defined in (\ref{coshshape}) with $a/r_0=6$, as a function of the spatial coordinate $z$, and the normalized channel index, $j/k_Fr_{min}$. In panels (c,d) $\ell_{ee}/r_0=0.3$. Green color corresponds to zero population, while white reflects states above the Fermi energy.} 
\end{figure}

{\it Ballistic regime.} In the absence of collisions (${\tilde I}=0$), Eq. (\ref{Boltzmanng1}) implies that $h^j_{R,L}$ is independent of $z$ and is such that $h^j_R=h^j_L$ at the classical turning point, where $j=k_F r(z)$. The solution states that there is no inter-channel scattering along the wormhole, which is a consequence of angular momentum conservation. As for intra-channel back-scattering, two situations may exist: fully transmitted channels are those with $j<k_Fr_{min}$. For these channels, each of the two non-equilibrium occupations $h^j_{R,L}(z)$ is determined by the contact from which it emanates, and is independent of $z$. In contrast, if there is a point $z_0$ for which $j=k_F r(z_0)$, at this point the momentum has no $x$-component, $h_R=h_L$ and the channel is fully reflected. 
Then,  on one side of $z_0$ where the channel exists we have $h_R=h_L$, with the value being determined by the contact from which the channel emanates and to which it is back-reflected. Both occupations vanish at the other side of $z_0$, in which the channel does not exist. Figure (\ref{coshresults}a,b) presents $h_R^j\mp h_L^j$ for a particular example of a ballistic wormhole, showing the non-equilibrium channel-dependent contributions to the local potential and current density. 

Each contact feeds into the wormhole all channels below its potential, $\pm V_0/2$ for the left and right contacts respectively, thus specifying the boundary conditions. By Landauer's formula, $V_0=\pi I/k_Fr_{min}$. With these boundary conditions, we can solve for $h^j_{R,L}(z)$ and use the solution to calculate the potential as a function of $z$. We find the potential to be, 
\begin{eqnarray}
    V_{\rm ballistic}(z)&=&-\sgn (z) \frac{V_0}{\pi}\int_{k_Fr_{min}}^{k_Fr(z)}\frac{dj}{\sqrt{(k_Fr(z))^2-j^2}}\nonumber\\ 
    &=&-\sgn z \frac{V_0}{2}\left[1-\frac{2}{\pi}\arcsin{\frac{r_{min}}{r(z)}}\right]
    \label{eq:potentialballistic}
\end{eqnarray}

Interestingly, although there are no collisions, we see that there is a potential drop, and thus resistance, in the bulk of the wormhole. Eq. (\ref{eq:potentialballistic}) shows that the potential close to the edges of the wormhole ($z=\pm L/2$) is smaller than that in the contacts  by $\frac{V_0}{\pi}\arcsin{\frac{r_{min}}{r_{max}}}$. This difference is the LS contact resistance. In the limit $r_{max}\gg r_{min}$ this contact resistance becomes negligible, and practically all the LS resistance drops  in the bulk. From Eq. (\ref{eq:potentialballistic}) we see that voltage drops in the bulk when the upper limit of the integral varies with $z$. Hence,  the bulk LS resistance appears whenever the number of conduction channels varies in the bulk. As we show below, electron-electron scattering can dramatically suppress the bulk potential drop, allowing the system to conduct much better than the fundamental LS limit.

{\it Electron-electron scattering and the hydrodynamic regime.} We now turn to consider the effect of momentum conserving electron-electron interactions on the wormhole resistance. Within the relaxation time approximation, taking conservation laws into account \cite{Callaway,PhysRevB.51.13389}, we have
\begin{equation}
I[f({\bf p, r})]=-\frac{1}{\ell_{ee}}\left [f-\frac{\delta(\epsilon_F-\epsilon({\bf p}))}{\nu_F} (\rho({\bf r})+\frac{2J_x({\bf r})\cos\theta}{ v_F})\right ]
\end{equation}
The second and third terms on the right hand side guarantee charge and momentum conservation, respectively. We obtain ${\tilde I}[h^j]$ using the same ansatz we used before, 
\begin{equation}
\begin{split}
     &{\tilde I}[{h^j_{R,L}}({z})] = \\ &-\frac{\nu^j}{\ell_{ee}}\left [h^j_{R,L}(z) -\frac{\rho(z)}{\nu_F}\mp
  \frac{4\pi J_x(z)}{k_F}\sqrt{1-\left (\frac{j}{k_Fr(z)}\right)^2}\right ]
      \label{scatteringtilde}
  \end{split}
 \end{equation}
 The $\nu^j/\ell_{ee}$ factor makes the mean free path $j$-dependent and shortens it from $\ell_{ee}$ to $\ell_{ee}\sqrt{1-\left (\frac{j}{k_Fr(z)}\right)^2}$. This may be understood by noting that for $j/r(z)$ large, $p_x$ is small and a shorter distance is traversed in the $z$-direction between two scattering events. In particular, the scattering length vanishes when the channel is about to be terminated, opening a way for the electrons to avoid back-scattering by being scattered to a transmitted channel. Furthermore, in contrast to the case of impurity scattering, in which in Eq. (\ref{scatteringtilde}) $\ell_{ee}$ is replaced by a momentum-relaxing mean free path and the third term is absent, here the presence of the third term allows for a Galilean boost of the Fermi surface ${h^j}_{R,L}(z)=\pm\frac{4\pi J_x({z})}{k_F}\sqrt{1-(j/k_Fr(z))^2}$ to be carried out without developing  a resistance. 

We find the solution to a leading order in $\ell_{ee}$ (the calculation is given in the Appendix), 
\begin{eqnarray}
    h^j_{R,L}(z)&=&\pm\frac{2I}{ k_F r(z)}\sqrt{1-\left (\frac{j}{k_Fr(z)}\right)^2} \nonumber \\
    &+&\frac{2I\ell_{ee}\sin\xi(z)}{ k_F r^{2}(z)} [1-2 (\frac{j}{k_Fr(z)})^2 ] \nonumber \\
    &-&\int_0^z \frac{I\ell_{ee}}{ k_{F}r^{2}(z')}\cos\xi(z')\frac{d\xi}{dz'}(z')dz'\label{eq:wormholesolution1}
\end{eqnarray}
where $\xi(z)$ is the local angle between $z$-axis and the manifold, i.e. $r(z)'=\tan{\xi(z)}$. This solution is valid in the bulk, away from the contacts. We comment on the role of the contacts below, with details in the Appendix.

The first term in Eq. (\ref{eq:wormholesolution1}) is a rigidly shifted Fermi surface. It is the  solution expected for $r'=0$ far away from the contacts, after all deformations of the Fermi surface are suppressed by the scattering term. The second and third terms are smaller than the first by a factor of $\ell_{ee}/r(z)$, and  originate from the breaking of Galilean invariance. The second term makes the shifted Fermi surface acquire an elongated shape, with more electrons in the head-on direction (small $j$), and less in the $j\approx k_Fr(z)$ channels. The third term is independent of $j$. It carries an electronic density, and  leads to a potential drop and  resistivity. Note that while the second term exists when $\sin\xi\ne 0$, the third term requires $\frac{d\sin\xi}{dz}\ne 0$. Stated differently, in contrast to ballistic electrons for which local resistance appears when the number of conduction channels varies with $z$, i.e., when $r'\ne 0$, for hydrodynamic electrons resistance is generated only when this function has a non-zero curvature, $r''\ne 0$. 

The potential originating from the third term of Eq. (\ref{eq:wormholesolution1}) may be written also as:
\begin{equation}
 V_{\rm hydro}(z)=
 I\int_0^{\xi(z)}\frac{\ell_{ee}}{4\pi k_{F}r^{2}(\xi)}\cos\xi d\xi \label{potential}   
\end{equation}

 The  resistance scale may be estimated from Eq. (\ref{potential}). The $r^2$ in the denominator suggests that the wormhole resistance is characterized by a ``super-ballistic'' scale \cite{Guo3068,Kumar2017,PhysRevLett.123.026801,Hong2020}, $\frac{2\pi\ell_{ee}}{k_F r_{min}^2}$, smaller by $2\ell_{ee}/r_{min}$ than the ballistic LS resistance. However, Eq. (\ref{potential}) opens the way for a much smaller scale, $\frac{\ell_{ee}}{4\pi k_F r_{min}^2}\sin\xi_0$, where $\xi_0$ is the angle at which $r$ becomes much larger than $r_{min}$. If $r$ grows slowly, $\sin\xi_0$ may be much smaller than one, with the resistance becoming much smaller than the super-ballistic scale. Consequently, for a fixed $r_{max}\gg r_{min}$ the resistance generally decreases with increasing $L$, opposite to the familiar Ohmic dependence. 
 
\begin{figure}[t]
\centering
\includegraphics[width=.4\textwidth,scale=0.05]{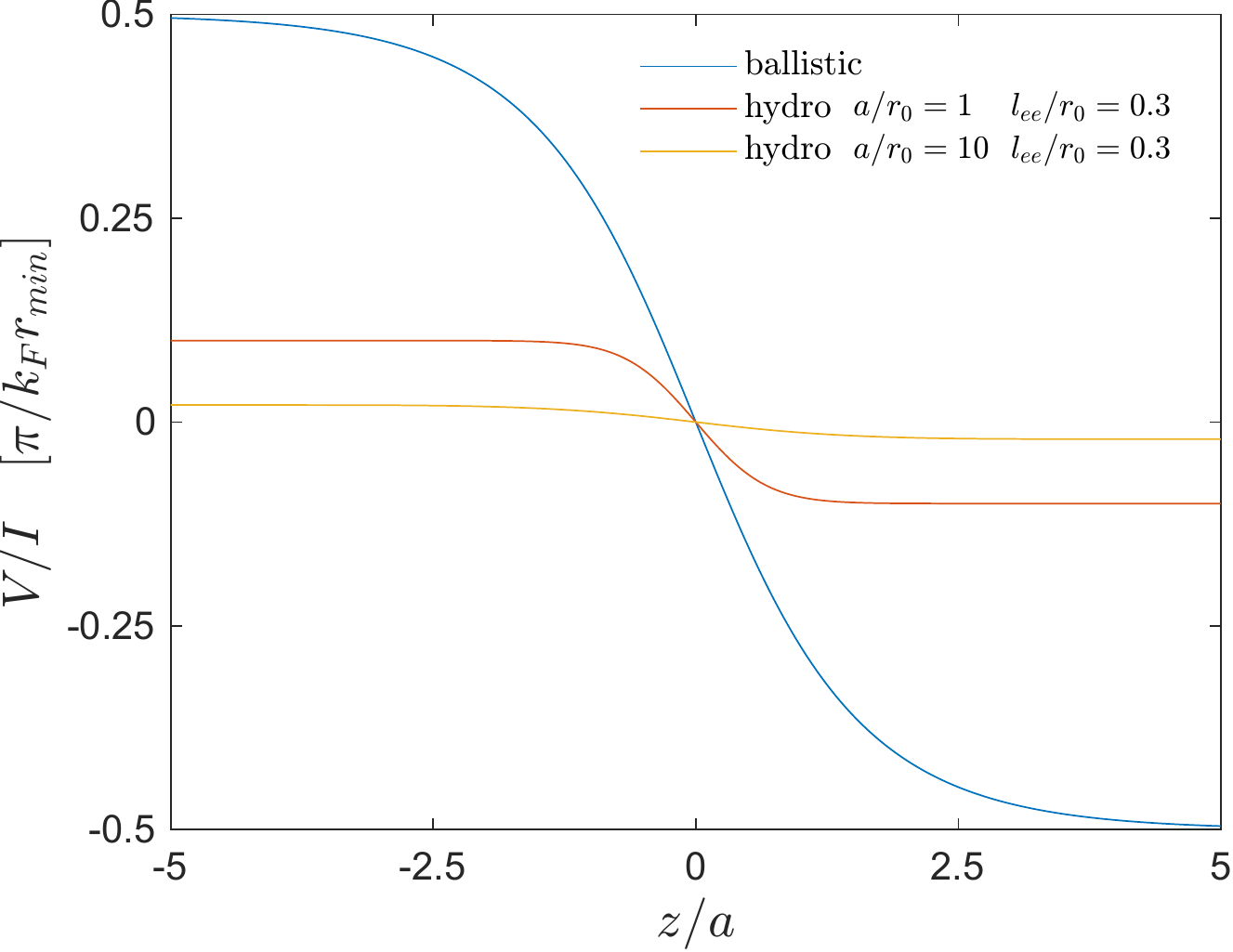}
\caption{\label{cosh-potential-dist} The potential along the wormhole defined in (\ref{coshshape}), divided by the current, $V/I$, in units of the LS resistance, plotted for a ballistic flow ($\ell_{ee}=\infty$) and hydrodynamic flows ($\ell_{ee}/r_{min}=0.3$) with varying values of $a/r_0$ (see legend).} 
\end{figure} 
 
To illustrate the two hydrodynamic scales, consider an example where 
 \begin{equation}
     r(z)=r_0\cosh{z/a}
     \label{coshshape}
 \end{equation}
 In this wormhole  $r_{min}=r_0$ and $r_{max}\gg r_{min}$ for $L\gg a$. Under the latter condition, the contribution to the resistance decays fast with $|z|\gg a$, and we can take $L\rightarrow\infty$. Then, 
using Eq. (\ref{potential}),
\begin{equation}
    R_{\cosh}=\frac{\ell_{ee}}{2\pi k_F}\left [\frac{1}{r_0^2-a^2}-\frac{a^2 \arctanh{\frac{\sqrt{r_0^2-a^2 }}{r_0 }}}{r_0 (r_0^2-a^2)^{3/2}}\right ]
\end{equation}
In the limit $a\rightarrow 0$ the resistance tends to $\frac{\ell_{ee}}{2\pi k_Fr_0^2}$, but when $a\gg r_0$  it decreases to become of order $\frac{\ell_{ee}}{4 k_Fr_0a}$. As can be seen in Eq. (\ref{potential}), most of the resistance originates from the product of the minimum radius $r_{min}$ and the change in angle $\Delta\xi$ over which the radius becomes significantly larger than $r_0$. When $a\gg r_0$ the change in angle is $\Delta\xi\sim r_0/a$ and hence the decrease in resistance. Fig. (\ref{coshresults}c,d) show the calculated $h_L-h_R$ and $h_L+h_R$ for hydrodynamic flow in the wormhole in Eq. (\ref{coshshape}). These quantities contribute to the current density and potential, respectively (Eq. (\ref{currentdensity})). Fig. (\ref{cosh-potential-dist}) shows potential drop in this wormhole as a function of $z$, in the ballistic case and in the hydrodynamic cases for two values of $a/r_0$. The hydrodynamic suppression of the resistance with increasing $a$ is evident. Note that when  $r(z)$ is constant the resistance in the bulk vanishes, since the bulk is Galilean invariant. However, the LS voltage drop occurs then sharply at the contacts, and is not suppressed by electron-electron scattering. To suppress the resistance by electron-electron scattering $r(z)$ should vary slowly from $r_{min}$ to  $r_{max}\gg r_{min}$.

Our analysis elucidates this suppression of the resistance: a potential drop results from reflection of electrons. In the ballistic regime the contact sends into the sample electrons in channels for which $j$ is too large to be transmitted. Those electrons are reflected, leading to a voltage drop (Eq. (\ref{eq:potentialballistic})). In contrast, in the hydrodynamic regime electrons of high $j$ are scattered to channels of smaller $j$, and largely end up being transmitted, without generating a potential drop. Note that our entire analysis assumes $\ell_{ee}\ll a$ and $\ell_{ee}\ll r_0$, in contrast to the sharp constriction case, studied, e.g., in  \onlinecite{Guo3068}, leading to a rather different evolution of $R$ with $\ell_{ee}$. 
 
Eqs. (\ref{potential}) and (\ref{coshshape})  show that the bulk resistance of a Corbino disk vanishes, as a consequence of the lack of variation of $\xi$. With the limitation of $z$ to a proper range, and with the limit $a\rightarrow 0$, Eq. (\ref{coshshape}) may be used to describe a Corbino disk.   The resulting bulk resistance vanishes in that limit. Indeed, in a Corbino disk the number of channels grows linearly with the radial coordinate, its second derivative vanishes, and so does the hydrodynamic resistance. Importantly, this vanishing bulk resistance is in series with a contact resistance which in this case is $\pi/(2 k_F r_{min})$, where $r_{min}$ is the inner radius of the disk. 

The elimination of the LS resistance in a Corbino disk was experimentally confirmed, as reported in a companion article\cite{companion}. In that article, we generalized the present calculations to include momentum relaxation due to phonon and impurity scattering, and showed that it leads to a simple additive contribution to the resistance.

Finally, although the wormhole is illuminating theoretically, it is a rather exotic geometry for real-life transistor devices. Those typically have a  long rectangular bar geometry, in which the density varies along the $x$-axis and is maximal near the contacts. In a bar geometry, previous work (e.g. \cite{Moll1061}) has focused on a viscous contribution arising from the no-slip boundary condition. Here, we neglect this contribution by assuming specular boundary scattering, or a wide bar. 
By carrying out an analysis similar to that of the wormhole (see Appendix), we find the resistance 
\begin{equation}
    R_{bar}=\int_{-\infty}^\infty dx\frac{(k_F'\ell_{ee})'}{2k_{F}^2r}.
    \label{varyingdensitybar}
\end{equation}
Here we account also for the possibility that $\ell_{ee}$ varies with the variation of $k_F$. 
Assuming that the change in $k_F$ is much larger than its minimnal value, we can estimate $R\sim \ell_{ee}/k_Fra$, where $a$ is the scale over which $k_F$ and $\ell_{ee}$ become much larger than their minimal value.

In summary, we showed here that when the LS resistance of an electronic system is spread into its bulk, rather than being localized at the interface with the contacts, it may be significantly reduced by electron-electron scattering, in principle all the way down to zero. 


\begin{acknowledgements}
We thank L. Ella, G. Falkovich, L. Levitov, M. Polini,
M. Shavit, A. Rozen, A. V. Shytov and U. Zondiner for
useful discussions. Work was supported by the Leona M.
and Harry B. Helmsley Charitable Trust grant, ISF grant
(No. 1182/21), Minerva grant (No. 713237), Hydrotronics
(No. 873028) and the ERC-Cog (See-1D-Qmatter,
No. 647413). T. S. acknowledges the support of the
Natural Sciences and Engineering Research Council of
Canada (NSERC), in particular the Discovery Grant
(No. RGPIN-2020-05842), the Accelerator Supplement
(No. RGPAS-2020-00060) and the Discovery Launch
Supplement (No. DGECR-2020-00222). This research
was enabled in part by support provided by Compute
Canada. A. S. was supported by the ERC under the Horizon
2020 Research and Innovation programme (LEGOTOP
No. 788715), the DFG (CRC/Transregio 183, EI 519/7-1),
and the ISF Quantum Science and Technology (2074/19).
\end{acknowledgements}

\bibliography{bibliography}

\appendix
\onecolumngrid
\setcounter{secnumdepth}{3}
\setcounter{figure}{0}
\setcounter{equation}{0}
\newpage\clearpage
\input{SM}

\end{document}

%% file: SM.tex
\setlength{\belowcaptionskip}{0pt}

\begin{center}
    \large\textbf{Supplemental material to ``\textit{\mytitle}''}
\end{center}

In this supplemental material, we give (A) the derivation of the Boltzmann equation in the wormhole geometry, (B) the solution of the Boltzmann equation in the hydrodynamic regime for a wormhole, (C) the solution of the Boltzmann equation in the hydrodynamic regime for a bar with a density variation, and (D) a discussion of the contact resistance.

\section{Derivation of Boltzmann equation for the wormhole}

The wormhole is a surface of revolution defined, in a cylindrical system of coordinates,  by $r(z)$. In Cartesian components it is defined $(r(z)\cos(\phi),r(z)\sin(\phi),z)$.
The kinetic energy of a particle (which is also the Lagrangian) is given by 
\bea
|v_F|^2 = (r \dot{\phi})^2 + \dot{z}^2 (1 + r'^2)
\eea
and is conserved.
Since the norm of the velocity is conserved, we only need to keep track of its angle $\theta$, measured with respect to an arbitrary axis. We define $\theta$ by  :
\bea
r \dot{\phi} &= v_F \sin(\theta) \\
\sqrt{1 + r'^2} \dot{z} &= v_F \cos(\theta)
\eea

The equations of motion are given by:
\bea
\ddot{\phi} &= - 2 \frac{r'}{r} \dot{z} \dot{\phi} \\
\ddot{z} &= \frac{r r'}{1+r'^2} \dot{\phi}^2 - \frac{r' r''}{1 + r'^2} \dot{z}^2
\eea

We start from the initial 2D problem with 2 coordinates and 2 velocities. Time independent distribution of electrons will necessarily be rotational invariant, and hence independent of $\phi$. Furthermore, in the absence of a driving force the magnitudes of the momentum and the velocity are conserved, such that their dynamical variable is the he angle $\theta$, measured with respect to the $x$-axis. 
The Boltzmann equation is then an equation for two coordinates - the radial coordinate $r$ and the direction of the velocity $\theta$. The equations of motion for $r,\theta$:
\bea
\dot{r} &= r' \dot{z} = \frac{r'}{\sqrt{1+r'^2}} v_F \cos(\theta) \\
\dot{\theta} &= - \frac{r'}{r} \frac1{\sqrt{1+r'^2}} v_F \sin(\theta)
\eea

We can now write the Boltzmann equation for $f(r,\theta)$ as:
\bea
\frac{df}{dt} = \partial_r(f) \dot{r} + \partial_\theta (f) \dot\theta = v_F I[f]
\eea
with $I[f]$ the scattering integral given in the main text.
This leads to
\bea
\partial_r(f) \frac{r'}{\sqrt{1+r'^2}}  \cos(\theta) - \partial_\theta (f) \frac{r'}{r} \frac1{\sqrt{1+r'^2}}  \sin(\theta) =  I[f]
\eea

This can be rewritten as
\bea
\partial_r(f)   \cos(\theta) - \partial_\theta (f) \frac{1}{r}   \sin(\theta) = \frac{\sqrt{1+r'^2}}{r'}  I[f]
\label{Boltzmann-app-theta}
\eea
The left hand side is independent of $z$, which means in the ballistic regime, all surfaces are equivalent once expressed in terms of $r$.
In the non-ballistic case, the only difference between surfaces is that the scattering rate acquires a dependence on $r'$ through the factor $\sqrt{1+r'^2}/r'$.
Multiplying Eq. (\ref{Boltzmann-app-theta}) by $r'$, and using $r'\partial_r=\partial_z$ we obtain Eq. (\ref{Boltzmann}). 

\section{Wormhole and Corbino disk - solution of the Boltzmann equation in the hydrodynamic regime}
The equation describing the non-equilibrium current distribution is obtained from Eq. (\ref{Boltzmann-app-theta}) by setting $f({\bf p},z)=\delta(\epsilon_F-\epsilon({\bf p}))h^y(p_y)$ and integrating $p_F\int \frac{d p_x}{2\pi}$. We get,  
\begin{equation}
    \pm\left[{ \nu_F}\partial_z {h^y}_{R,L} - {\nu_F}\frac{r'}{r} p_y \partial_{p_y} {h^y}_{R,L}\right ]= \sqrt{1+r'^2} {\tilde I}[{h^y}]
    \label{Boltzmanng}
\end{equation}
with, 
\begin{equation}
  {\tilde I}[{h^y_{R,L}}({\bf p, r})]=-\frac{\nu^y}{\ell_{ee}}\left [{h^y}_{R,L} -\frac{\rho({\bf r})}{2\nu_F}\mp\frac{4\pi j_x({\bf r})}{k_F}\sqrt{1-(p_y/k_F)^2}\right ]
  \label{scatteringtilde2}
\end{equation}
The transformation of this equation to an equation for $h^j$ is explained below.
\subsection{Deriving the Boltzmann equation for $h^j$}

To get the Boltzmann equation that appears in Eq.~(\ref{Boltzmanng1}) of the main text, we start from Eq. (\ref{Boltzmanng}) and change the variables $z,p_{y}\to z,j$. Since $j=p_{y}r\left(z\right)$ we must treat the dependence between the coordinates carefully. It is easy to confirm that the second term in the LHS of (\ref{Boltzmanng}) becomes $-j\frac{r'}{r}\partial_{j}h_{R,L}$, however, the first term also changes because it is a total derivative, $\partial_{z}h_{R,L}\to\partial_{z}h_{R,L}+\frac{\partial j}{\partial z}\partial_{j}h_{R,L}$. Using $\frac{\partial j}{\partial z}=p_{y}r'$ we get that $$\frac{\partial j}{\partial z}\partial_{j}h_{R,L}-j\frac{r'}{r}\partial_{j}h_{R,L}=0,$$ thus obtaining Eq. (\ref{Boltzmanng1}). 

\subsection{Solving the Boltzmann equation to linear order in $\ell_{ee}$}
In this subsection we solve the equation for $h^j$ up to first order in $\ell_{ee} r'/r$.   

Assuming that $r(z)$ varies slowly on a scale of $\ell_{ee}$ we try as a first attempt the $z$-independent solution of a uniform $r$, adjusted at each point to the local $r(z)$, i.e., $\pm\frac{2I}{k_{F}r\left(z\right)}\sqrt{1-\left(\frac{j}{k_{F}r\left(z\right)}\right)^{2}}$. This attempted solution conserves current. Substituting it  into Eq. (\ref{Boltzmanng1}) we find on the LHS a remainder term $-\frac{2I}{k_{F}}\frac{r'\left(1-2\left(\frac{j}{k_{F}r}\right)^{2}\right)}{r^{2}\sqrt{1-\left(\frac{j}{k_{F}r\left(z\right)}\right)^{2}}}$. It is linear in $r'$, as expected. To compensate for this term, we  modify our solution, making it
\begin{eqnarray}
h_{R,L}^{j}\left(z\right)&=&\pm\frac{2I}{k_{F}r\left(z\right)}\sqrt{1-\left(\frac{j}{k_{F}r\left(z\right)}\right)^{2}}+\frac{2I\ell_{ee}r'}{k_{F}r^{2}\left(z\right)\sqrt{1+r'^{2}}}\left[1-2\left(\frac{j}{k_{F}r\left(z\right)}\right)^{2}\right]\nonumber \\
    &=&\pm\frac{2I}{k_{F}r\left(z\right)}\sqrt{1-\left(\frac{j}{k_{F}r\left(z\right)}\right)^{2}}+\frac{2I\ell_{ee}\sin\xi\left(z\right)}{k_{F}r^{2}\left(z\right)}\left[1-2\left(\frac{j}{k_{F}r\left(z\right)}\right)^{2}\right]
    \label{h2}
\end{eqnarray}
The term we added to $h$ is linear in $\ell_{ee}$, such that when substituted into the RHS, it balances the remainder term on the LHS, which is $\ell_{ee}$-independent.  However, it generates a new remainder term on the LHS, which is linear in $\ell_{ee}$. Specifically, this term is, 
\begin{equation}
    \pm\frac{2I\ell_{ee}}{k_{F}}\frac{r\xi'\cos\xi\left(1-2\left(\frac{j}{k_{F}r}\right)^{2}\right)+2r'\sin\xi\left(4\left(\frac{j}{k_{F}r}\right)^{2}-1\right)}{r^{3}}
    \label{remainderfix}
\end{equation}
Naively, we should balance this term by adding a term $\delta h\propto \ell_{ee}^2$ to our solution, thereby generating a term $\delta h\nu^j/\ell_{ee}$ on the RHS to cancel the contribution (\ref{remainderfix}) on the LHS. We should note, however, that the RHS cannot cancel parts of $\delta h$ that carry a current or a density. Furthermore, we cannot add to $h$ a term that carries current, because we assume a fixed driven current. As it turns out, the second term in (\ref{remainderfix}) can be cancelled by the RHS, but the first term requires more care.  We write it as 
\begin{equation}
    \pm\frac{I\ell_{ee}}{k_{F}r^{2}}\xi'\cos\xi\pm\frac{I\ell_{ee}}{k_{F}r^{2}}\xi'\cos\xi\left(1-4\left(\frac{j}{k_{F}r}\right)^{2}\right)
\end{equation}
and cancel the first, $j$-independent, term by subtracting a $j$-independent, density carrying, term, 
\begin{equation}
    \int dz'\frac{I\ell_{ee}}{k_{F}r^{2}}\xi'\cos\xi
\end{equation}
When divided by $\nu_F$, this term gives the local electrochemical potential.

\section{Bar with a density variation - solution of the Boltzmann equation for the hydrodynamic regime}

We consider an infinite bar parallel to the z-axis, in which the equilibrium density, and hence $k_F$ vary with $z$. Furthermore, with a variation of density comes also a variation of $\ell_{ee}$. When the walls of the bar are specular, we can view it as a cylinder,  and we denote the circumference by $2\pi r$. With these assumptions, Boltzmann equation (\ref{Boltzmanng1}) and the collision term (\ref{scatteringtilde}) remain the same as they were for a wormhole, with $r'=0$:
\begin{equation}
    \pm\frac{\nu_{F}}{\nu_{j}}\partial_{z}h_{R,L}=-\frac{1}{\ell_{ee}}\left[h_{R,L}-\frac{\rho}{2\nu_{F}}\mp\frac{4\pi j_{x}}{k_{F}}\sqrt{1-\left(\frac{j}{k_{F}r}\right)^{2}}\right]
    \label{boltzmannBar}
\end{equation}
For the limit of small $\ell_{ee}$, we first try as a naive $\ell_{ee}$-independent  solution the locally Galilean boosted Fermi sphere: $h_{R,L}=\pm\frac{2I}{k_{F}r}\sqrt{1-\left(\frac{j}{k_{F}r}\right)^{2}}$, and we aim to find all corrections of order $\ell_{ee}$ to this term. Any amendment we do to the naive solution  should not carry current, since the current is fixed to $I$. When the naive solution  is substituted in (\ref{boltzmannBar})  the right hand side vanishes since $I=2\pi rj_{x}$ and $\rho=0$. However we get an extra term of $-\frac{2Ik_{F}'}{k_{F}^{2}r}\left(1-2\left(\frac{j}{k_{F}r}\right)^{2}\right)$ on the LHS. We can balance this term by adding to our solution $\frac{2I\ell_{ee} k_{F}'}{k_{F}^{2}r}\left(1-2\left(\frac{j}{k_{F}r}\right)^{2}\right)$. 

This amended solution solves (\ref{boltzmannBar}), up to a remainder term on the LHS:
\begin{eqnarray}
  \sqrt{1-\left(\frac{j}{k_{F}r}\right)^{2}} \partial_z\left [ \frac{2I\ell_{ee} k_{F}'}{k_{F}^{2}r}\left(1-2\left(\frac{j}{k_{F}r}\right)^{2}\right)\right ]
  \label{remainder2}
\end{eqnarray}
This remainder term is of the order $\ell_{ee}$ and we need to amend our solution further to eliminate it. In principle, there are two ways to do that. The part of (\ref{remainder2}) that does not carry density or current can be eliminated by an addition of a term of order $\ell_{ee}^2$ to $h$. Such a term will yield an order $\ell_{ee}$ term on  the RHS. However, being a contribution to $h$ that is of order $\ell_{ee}^2$, it is beyond our scope. The part of (\ref{remainder2}) that carries density of current, on the other hand, should be canceled by adding a term of order $\ell_{ee}$ to $h$, that is purely a density term. Such a term will not affect the RHS, and its substitution in the LHS will cancel the terms in (\ref{remainder2}). It is this term we are after. 
An inspection of (\ref{boltzmannBar}) and (\ref{remainder2}) allows us to find it and write the full expression of $h$ to linear order in $\ell_{ee}$ as 
\begin{eqnarray}
    h^j_{R,L}&=&\pm\frac{2I}{k_{F}r}\sqrt{1-\left(\frac{j}{k_{F}r}\right)^{2}}\nonumber \\ 
    &+&\frac{2I\ell_{ee}}{k_{F}^{2}r}k_{F}'\left(1-2\left(\frac{j}{k_{F}r}\right)^{2}\right)
    \nonumber \\
    &-&I\int^z d\tilde{z}\frac{\left(k_{F}'\ell_{ee}\right)'}{2k_{F}^{2}r}
\end{eqnarray}
where in the integrand in the last term $k_F,\ell_{ee}$ are both functions of  $\tilde z$.

\section{Contact resistance}

At the contact the density variation is fast. As a consequence the full solution of the Boltzmann equation becomes hard to obtain, but we can still estimate the voltage drop on the contact region. For the simplest case of a ballistic  cylinder ($\xi=0$), or a ballistic rectangular-shaped conductor with specular walls, the entire potential drop is on the two contacts. By symmetry, at the center of the wormhole $h^j_R=-h^j_L=\frac{I}{2k_F r}$ and consequently $V=0$. In fact, these values of $h^j_R,h^j_L$ hold anywhere within the sample, at $|z|<L/2$. In the contacts ($z=\pm(L/2+\epsilon))$ themselves $h^j_R=h^j_L$, and their value is determined by the local potential. Thus, there is a jump in the value of $h^j_L$ at $z=-L/2$ and of $h^j_R$ at $z=L/2$, and this jump leads to the expected jump of the potentials at the contacts. 

Next, we think of the cylindrical geometry with electron-electron scattering. Far from the contacts (a distance much larger than $\ell_{ee}$) our main-text analysis holds, leading to  $h^j_R=-h^j_L=\frac{4\pi J_x}{k_F}\sqrt{1-(j/k_Fr)^2}$ and $V=0$. At the two contacts  $h^j_R=h^j_L=\pm V(I)/2$, and it is $V(I)$ that we estimate now. For clarity we focus on the left contact, at $z=-L/2$. 

The equations satisfied by $h_{R,L}^j$ are of the form
\begin{eqnarray}
    \pm{ \nu_F}\partial_z h^j_{R,L}(j, z)= -\frac{\nu^j}{\ell_{ee}}\left [{h^j}_{R,L} -\frac{\rho(z)}{\nu_F}\mp 
  \frac{4\pi J_x}{k_F}\sqrt{1-\left (\frac{j}{k_Fr}\right)^2}\right ]
    \label{Boltzmanng2}
\end{eqnarray}

 These equations are equivalent to, 
\begin{eqnarray}
    h_{R,L}^j=h_{R,L}^j(\pm L/2)e^{ \frac{\pm(z\pm L/2)\nu_j}{\nu_F\ell_{ee}}}  
    +\int_{z_i}^z &dz'& \frac{\nu_j}{\nu_F\ell_{ee}} \left [-\frac{\rho(z')}{\nu_F}\mp 
  \frac{4\pi J_x}{k_F}\right ]e^{ \frac{\pm(z\pm z')\nu_j}{\nu_F\ell_{ee}}}
    \label{schematicsolution}
\end{eqnarray}
Here, $h_{R,L}^j(\pm L/2)$ are the boundary conditions for the right- and left- moving electrons at the points where they enter the sample. The dependence of the distribution functions $h_{R,L}^j$ on $z$ near the left contact is very different for the left- and right- moving electrons. For the left-moving electrons the entry point is very far from the contact we look at, and therefore the initial condition is long forgotten.  In the bulk, the distribution of the left moving electrons does not vary in space, and since the second term in Eq.~(\ref{schematicsolution}) averages over a scale of $\ell_{ee}$, we expect the variation of $h^j_L$ to be slow even close to the contact. 

For the right-moving electrons, in contrast, near the contact  the solution is dominated by initial conditions. The distribution function, that starts as a constant $h^j_R=\pi^2 J_x/k_F$, decays into $h^j_R=\frac{4\pi J_x}{k_F}\sqrt{1-(j/k_F r)^2}$, at a distance  $\ell_{ee}/\nu_j$ from the contact. 

Motivated by these considerations, we make the ansatz, 
\begin{equation}
    \begin{split}
    h_L^j&\approx -\frac{4\pi J_x}{k_F}\sqrt{1-(j/k_F r)^2} \\
    h_R^j&=\frac{\pi^2 J_x}{k_F}e^{-(z+L/2)\nu^j/\nu_F\ell_{ee}}  + (1-e^{-(z+L/2)\nu^j/\ell_{ee}\nu_F})\frac{4\pi J_x}{k_F}\sqrt{1-(j/k_F r)^2}
    \label{ansatzcontact}
    \end{split}
\end{equation}

Within this ansatz, the voltage difference between the contact itself and the bulk is the same as it is in the ballistic case, namely half of the Landauer-Sharvin voltage drop falls on each contact. The effect of the scattering term is limited to distributing this voltage drop from being at the interface itself to being spread on a scale of $\ell_{ee}$.

%% file: main.bbl
\begin{thebibliography}{55}%
\makeatletter
\providecommand \@ifxundefined [1]{%
 \@ifx{#1\undefined}
}%
\providecommand \@ifnum [1]{%
 \ifnum #1\expandafter \@firstoftwo
 \else \expandafter \@secondoftwo
 \fi
}%
\providecommand \@ifx [1]{%
 \ifx #1\expandafter \@firstoftwo
 \else \expandafter \@secondoftwo
 \fi
}%
\providecommand \natexlab [1]{#1}%
\providecommand \enquote  [1]{``#1''}%
\providecommand \bibnamefont  [1]{#1}%
\providecommand \bibfnamefont [1]{#1}%
\providecommand \citenamefont [1]{#1}%
\providecommand \href@noop [0]{\@secondoftwo}%
\providecommand \href [0]{\begingroup \@sanitize@url \@href}%
\providecommand \@href[1]{\@@startlink{#1}\@@href}%
\providecommand \@@href[1]{\endgroup#1\@@endlink}%
\providecommand \@sanitize@url [0]{\catcode `\\12\catcode `\$12\catcode
  `\&12\catcode `\#12\catcode `\^12\catcode `\_12\catcode `\%12\relax}%
\providecommand \@@startlink[1]{}%
\providecommand \@@endlink[0]{}%
\providecommand \url  [0]{\begingroup\@sanitize@url \@url }%
\providecommand \@url [1]{\endgroup\@href {#1}{\urlprefix }}%
\providecommand \urlprefix  [0]{URL }%
\providecommand \Eprint [0]{\href }%
\providecommand \doibase [0]{http://dx.doi.org/}%
\providecommand \selectlanguage [0]{\@gobble}%
\providecommand \bibinfo  [0]{\@secondoftwo}%
\providecommand \bibfield  [0]{\@secondoftwo}%
\providecommand \translation [1]{[#1]}%
\providecommand \BibitemOpen [0]{}%
\providecommand \bibitemStop [0]{}%
\providecommand \bibitemNoStop [0]{.\EOS\space}%
\providecommand \EOS [0]{\spacefactor3000\relax}%
\providecommand \BibitemShut  [1]{\csname bibitem#1\endcsname}%
\let\auto@bib@innerbib\@empty
\bibitem [{\citenamefont {Landauer}(1957)}]{Landauer1957}%
  \BibitemOpen
  \bibfield  {author} {\bibinfo {author} {\bibfnamefont {R.}~\bibnamefont
  {Landauer}},\ }\href {\doibase 10.1147/rd.13.0223} {\bibfield  {journal}
  {\bibinfo  {journal} {IBM Journal of Research and Development}\ }\textbf
  {\bibinfo {volume} {1}},\ \bibinfo {pages} {223} (\bibinfo {year}
  {1957})}\BibitemShut {NoStop}%
\bibitem [{\citenamefont {{Sharvin}}(1965)}]{Sharvin1965}%
  \BibitemOpen
  \bibfield  {author} {\bibinfo {author} {\bibfnamefont {Y.~V.}\ \bibnamefont
  {{Sharvin}}},\ }\href@noop {} {\bibfield  {journal} {\bibinfo  {journal}
  {Soviet Journal of Experimental and Theoretical Physics}\ }\textbf {\bibinfo
  {volume} {21}},\ \bibinfo {pages} {655} (\bibinfo {year} {1965})}\BibitemShut
  {NoStop}%
\bibitem [{\citenamefont {Datta}(1995)}]{datta_1995}%
  \BibitemOpen
  \bibfield  {author} {\bibinfo {author} {\bibfnamefont {S.}~\bibnamefont
  {Datta}},\ }\href@noop {} {\emph {\bibinfo {title} {Electronic Transport in
  Mesoscopic Systems}}},\ Cambridge Studies in Semiconductor Physics and
  Microelectronic Engineering\ (\bibinfo  {publisher} {Cambridge University
  Press},\ \bibinfo {year} {1995})\BibitemShut {NoStop}%
\bibitem [{\citenamefont {Imry}(2002)}]{imry2002introduction}%
  \BibitemOpen
  \bibfield  {author} {\bibinfo {author} {\bibfnamefont {Y.}~\bibnamefont
  {Imry}},\ }\href@noop {} {\emph {\bibinfo {title} {Introduction to mesoscopic
  physics}}}\ (\bibinfo  {publisher} {Oxford University Press on Demand},\
  \bibinfo {year} {2002})\BibitemShut {NoStop}%
\bibitem [{\citenamefont {Gurzhi}(1963)}]{gurzhi1963minimum}%
  \BibitemOpen
  \bibfield  {author} {\bibinfo {author} {\bibfnamefont {R.}~\bibnamefont
  {Gurzhi}},\ }\href@noop {} {\bibfield  {journal} {\bibinfo  {journal} {Sov.
  Phys. JETP}\ }\textbf {\bibinfo {volume} {44}},\ \bibinfo {pages} {771}
  (\bibinfo {year} {1963})}\BibitemShut {NoStop}%
\bibitem [{\citenamefont {Gurzhi}(1968)}]{gurzhi1968hydrodynamic}%
  \BibitemOpen
  \bibfield  {author} {\bibinfo {author} {\bibfnamefont {R.}~\bibnamefont
  {Gurzhi}},\ }\href@noop {} {\bibfield  {journal} {\bibinfo  {journal}
  {Physics-Uspekhi}\ }\textbf {\bibinfo {volume} {11}},\ \bibinfo {pages} {255}
  (\bibinfo {year} {1968})}\BibitemShut {NoStop}%
\bibitem [{\citenamefont {Black}(1980)}]{PhysRevB.21.3279}%
  \BibitemOpen
  \bibfield  {author} {\bibinfo {author} {\bibfnamefont {J.~E.}\ \bibnamefont
  {Black}},\ }\href {\doibase 10.1103/PhysRevB.21.3279} {\bibfield  {journal}
  {\bibinfo  {journal} {Phys. Rev. B}\ }\textbf {\bibinfo {volume} {21}},\
  \bibinfo {pages} {3279} (\bibinfo {year} {1980})}\BibitemShut {NoStop}%
\bibitem [{\citenamefont {Yu}\ \emph {et~al.}(1984)\citenamefont {Yu},
  \citenamefont {Haerle}, \citenamefont {Zwart}, \citenamefont {Bass},
  \citenamefont {Pratt},\ and\ \citenamefont {Schroeder}}]{PhysRevLett.52.368}%
  \BibitemOpen
  \bibfield  {author} {\bibinfo {author} {\bibfnamefont {Z.~Z.}\ \bibnamefont
  {Yu}}, \bibinfo {author} {\bibfnamefont {M.}~\bibnamefont {Haerle}}, \bibinfo
  {author} {\bibfnamefont {J.~W.}\ \bibnamefont {Zwart}}, \bibinfo {author}
  {\bibfnamefont {J.}~\bibnamefont {Bass}}, \bibinfo {author} {\bibfnamefont
  {W.~P.}\ \bibnamefont {Pratt}}, \ and\ \bibinfo {author} {\bibfnamefont
  {P.~A.}\ \bibnamefont {Schroeder}},\ }\href {\doibase
  10.1103/PhysRevLett.52.368} {\bibfield  {journal} {\bibinfo  {journal} {Phys.
  Rev. Lett.}\ }\textbf {\bibinfo {volume} {52}},\ \bibinfo {pages} {368}
  (\bibinfo {year} {1984})}\BibitemShut {NoStop}%
\bibitem [{\citenamefont {Gurzhi}\ \emph {et~al.}(1995)\citenamefont {Gurzhi},
  \citenamefont {Kalinenko},\ and\ \citenamefont
  {Kopeliovich}}]{gurzhi1995electron}%
  \BibitemOpen
  \bibfield  {author} {\bibinfo {author} {\bibfnamefont {R.}~\bibnamefont
  {Gurzhi}}, \bibinfo {author} {\bibfnamefont {A.}~\bibnamefont {Kalinenko}}, \
  and\ \bibinfo {author} {\bibfnamefont {A.}~\bibnamefont {Kopeliovich}},\
  }\href@noop {} {\bibfield  {journal} {\bibinfo  {journal} {Physical review
  letters}\ }\textbf {\bibinfo {volume} {74}},\ \bibinfo {pages} {3872}
  (\bibinfo {year} {1995})}\BibitemShut {NoStop}%
\bibitem [{\citenamefont {Molenkamp}\ and\ \citenamefont
  {de~Jong}(1994)}]{PhysRevB.49.5038}%
  \BibitemOpen
  \bibfield  {author} {\bibinfo {author} {\bibfnamefont {L.~W.}\ \bibnamefont
  {Molenkamp}}\ and\ \bibinfo {author} {\bibfnamefont {M.~J.~M.}\ \bibnamefont
  {de~Jong}},\ }\href {\doibase 10.1103/PhysRevB.49.5038} {\bibfield  {journal}
  {\bibinfo  {journal} {Phys. Rev. B}\ }\textbf {\bibinfo {volume} {49}},\
  \bibinfo {pages} {5038} (\bibinfo {year} {1994})}\BibitemShut {NoStop}%
\bibitem [{\citenamefont {de~Jong}\ and\ \citenamefont
  {Molenkamp}(1995)}]{PhysRevB.51.13389}%
  \BibitemOpen
  \bibfield  {author} {\bibinfo {author} {\bibfnamefont {M.~J.~M.}\
  \bibnamefont {de~Jong}}\ and\ \bibinfo {author} {\bibfnamefont {L.~W.}\
  \bibnamefont {Molenkamp}},\ }\href {\doibase 10.1103/PhysRevB.51.13389}
  {\bibfield  {journal} {\bibinfo  {journal} {Phys. Rev. B}\ }\textbf {\bibinfo
  {volume} {51}},\ \bibinfo {pages} {13389} (\bibinfo {year}
  {1995})}\BibitemShut {NoStop}%
\bibitem [{\citenamefont {Dyakonov}\ and\ \citenamefont
  {Shur}(1993)}]{PhysRevLett.71.2465}%
  \BibitemOpen
  \bibfield  {author} {\bibinfo {author} {\bibfnamefont {M.}~\bibnamefont
  {Dyakonov}}\ and\ \bibinfo {author} {\bibfnamefont {M.}~\bibnamefont
  {Shur}},\ }\href {\doibase 10.1103/PhysRevLett.71.2465} {\bibfield  {journal}
  {\bibinfo  {journal} {Phys. Rev. Lett.}\ }\textbf {\bibinfo {volume} {71}},\
  \bibinfo {pages} {2465} (\bibinfo {year} {1993})}\BibitemShut {NoStop}%
\bibitem [{\citenamefont {Chow}\ \emph {et~al.}(1996)\citenamefont {Chow},
  \citenamefont {Wei}, \citenamefont {Girvin},\ and\ \citenamefont
  {Shayegan}}]{PhysRevLett.77.1143}%
  \BibitemOpen
  \bibfield  {author} {\bibinfo {author} {\bibfnamefont {E.}~\bibnamefont
  {Chow}}, \bibinfo {author} {\bibfnamefont {H.~P.}\ \bibnamefont {Wei}},
  \bibinfo {author} {\bibfnamefont {S.~M.}\ \bibnamefont {Girvin}}, \ and\
  \bibinfo {author} {\bibfnamefont {M.}~\bibnamefont {Shayegan}},\ }\href
  {\doibase 10.1103/PhysRevLett.77.1143} {\bibfield  {journal} {\bibinfo
  {journal} {Phys. Rev. Lett.}\ }\textbf {\bibinfo {volume} {77}},\ \bibinfo
  {pages} {1143} (\bibinfo {year} {1996})}\BibitemShut {NoStop}%
\bibitem [{\citenamefont {Spivak}\ and\ \citenamefont
  {Kivelson}(2006)}]{Spivak20062071}%
  \BibitemOpen
  \bibfield  {author} {\bibinfo {author} {\bibfnamefont {B.}~\bibnamefont
  {Spivak}}\ and\ \bibinfo {author} {\bibfnamefont {S.~A.}\ \bibnamefont
  {Kivelson}},\ }\href {\doibase http://dx.doi.org/10.1016/j.aop.2005.12.002}
  {\bibfield  {journal} {\bibinfo  {journal} {Annals of Physics}\ }\textbf
  {\bibinfo {volume} {321}},\ \bibinfo {pages} {2071 } (\bibinfo {year}
  {2006})}\BibitemShut {NoStop}%
\bibitem [{\citenamefont {Andreev}\ \emph {et~al.}(2011)\citenamefont
  {Andreev}, \citenamefont {Kivelson},\ and\ \citenamefont
  {Spivak}}]{PhysRevLett.106.256804}%
  \BibitemOpen
  \bibfield  {author} {\bibinfo {author} {\bibfnamefont {A.~V.}\ \bibnamefont
  {Andreev}}, \bibinfo {author} {\bibfnamefont {S.~A.}\ \bibnamefont
  {Kivelson}}, \ and\ \bibinfo {author} {\bibfnamefont {B.}~\bibnamefont
  {Spivak}},\ }\href {\doibase 10.1103/PhysRevLett.106.256804} {\bibfield
  {journal} {\bibinfo  {journal} {Phys. Rev. Lett.}\ }\textbf {\bibinfo
  {volume} {106}},\ \bibinfo {pages} {256804} (\bibinfo {year}
  {2011})}\BibitemShut {NoStop}%
\bibitem [{\citenamefont {Torre}\ \emph {et~al.}(2015)\citenamefont {Torre},
  \citenamefont {Tomadin}, \citenamefont {Geim},\ and\ \citenamefont
  {Polini}}]{PhysRevB.92.165433}%
  \BibitemOpen
  \bibfield  {author} {\bibinfo {author} {\bibfnamefont {I.}~\bibnamefont
  {Torre}}, \bibinfo {author} {\bibfnamefont {A.}~\bibnamefont {Tomadin}},
  \bibinfo {author} {\bibfnamefont {A.~K.}\ \bibnamefont {Geim}}, \ and\
  \bibinfo {author} {\bibfnamefont {M.}~\bibnamefont {Polini}},\ }\href
  {\doibase 10.1103/PhysRevB.92.165433} {\bibfield  {journal} {\bibinfo
  {journal} {Phys. Rev. B}\ }\textbf {\bibinfo {volume} {92}},\ \bibinfo
  {pages} {165433} (\bibinfo {year} {2015})}\BibitemShut {NoStop}%
\bibitem [{\citenamefont {Alekseev}(2016)}]{PhysRevLett.117.166601}%
  \BibitemOpen
  \bibfield  {author} {\bibinfo {author} {\bibfnamefont {P.~S.}\ \bibnamefont
  {Alekseev}},\ }\href {\doibase 10.1103/PhysRevLett.117.166601} {\bibfield
  {journal} {\bibinfo  {journal} {Phys. Rev. Lett.}\ }\textbf {\bibinfo
  {volume} {117}},\ \bibinfo {pages} {166601} (\bibinfo {year}
  {2016})}\BibitemShut {NoStop}%
\bibitem [{\citenamefont {Tomadin}\ \emph {et~al.}(2014)\citenamefont
  {Tomadin}, \citenamefont {Vignale},\ and\ \citenamefont
  {Polini}}]{PhysRevLett.113.235901}%
  \BibitemOpen
  \bibfield  {author} {\bibinfo {author} {\bibfnamefont {A.}~\bibnamefont
  {Tomadin}}, \bibinfo {author} {\bibfnamefont {G.}~\bibnamefont {Vignale}}, \
  and\ \bibinfo {author} {\bibfnamefont {M.}~\bibnamefont {Polini}},\ }\href
  {\doibase 10.1103/PhysRevLett.113.235901} {\bibfield  {journal} {\bibinfo
  {journal} {Phys. Rev. Lett.}\ }\textbf {\bibinfo {volume} {113}},\ \bibinfo
  {pages} {235901} (\bibinfo {year} {2014})}\BibitemShut {NoStop}%
\bibitem [{\citenamefont {Levitov}\ and\ \citenamefont
  {Falkovich}(2016)}]{levitov}%
  \BibitemOpen
  \bibfield  {author} {\bibinfo {author} {\bibfnamefont {L.}~\bibnamefont
  {Levitov}}\ and\ \bibinfo {author} {\bibfnamefont {G.}~\bibnamefont
  {Falkovich}},\ }\href {http://dx.doi.org/10.1038/nphys3667} {\bibfield
  {journal} {\bibinfo  {journal} {Nat Phys}\ }\textbf {\bibinfo {volume}
  {12}},\ \bibinfo {pages} {672} (\bibinfo {year} {2016})}\BibitemShut
  {NoStop}%
\bibitem [{\citenamefont {Principi}\ \emph {et~al.}(2016)\citenamefont
  {Principi}, \citenamefont {Vignale}, \citenamefont {Carrega},\ and\
  \citenamefont {Polini}}]{PhysRevB.93.125410}%
  \BibitemOpen
  \bibfield  {author} {\bibinfo {author} {\bibfnamefont {A.}~\bibnamefont
  {Principi}}, \bibinfo {author} {\bibfnamefont {G.}~\bibnamefont {Vignale}},
  \bibinfo {author} {\bibfnamefont {M.}~\bibnamefont {Carrega}}, \ and\
  \bibinfo {author} {\bibfnamefont {M.}~\bibnamefont {Polini}},\ }\href
  {\doibase 10.1103/PhysRevB.93.125410} {\bibfield  {journal} {\bibinfo
  {journal} {Phys. Rev. B}\ }\textbf {\bibinfo {volume} {93}},\ \bibinfo
  {pages} {125410} (\bibinfo {year} {2016})}\BibitemShut {NoStop}%
\bibitem [{\citenamefont {Sherafati}\ \emph {et~al.}(2016)\citenamefont
  {Sherafati}, \citenamefont {Principi},\ and\ \citenamefont
  {Vignale}}]{PhysRevB.94.125427}%
  \BibitemOpen
  \bibfield  {author} {\bibinfo {author} {\bibfnamefont {M.}~\bibnamefont
  {Sherafati}}, \bibinfo {author} {\bibfnamefont {A.}~\bibnamefont {Principi}},
  \ and\ \bibinfo {author} {\bibfnamefont {G.}~\bibnamefont {Vignale}},\ }\href
  {\doibase 10.1103/PhysRevB.94.125427} {\bibfield  {journal} {\bibinfo
  {journal} {Phys. Rev. B}\ }\textbf {\bibinfo {volume} {94}},\ \bibinfo
  {pages} {125427} (\bibinfo {year} {2016})}\BibitemShut {NoStop}%
\bibitem [{\citenamefont {{Guo}}\ \emph {et~al.}(2016)\citenamefont {{Guo}},
  \citenamefont {{Ilseven}}, \citenamefont {{Falkovich}},\ and\ \citenamefont
  {{Levitov}}}]{2016arXiv161209239G}%
  \BibitemOpen
  \bibfield  {author} {\bibinfo {author} {\bibfnamefont {H.}~\bibnamefont
  {{Guo}}}, \bibinfo {author} {\bibfnamefont {E.}~\bibnamefont {{Ilseven}}},
  \bibinfo {author} {\bibfnamefont {G.}~\bibnamefont {{Falkovich}}}, \ and\
  \bibinfo {author} {\bibfnamefont {L.}~\bibnamefont {{Levitov}}},\ }\href@noop
  {} {\bibfield  {journal} {\bibinfo  {journal} {ArXiv e-prints}\ } (\bibinfo
  {year} {2016})},\ \Eprint {http://arxiv.org/abs/1612.09239} {arXiv:1612.09239
  [cond-mat.mes-hall]} \BibitemShut {NoStop}%
\bibitem [{\citenamefont {Lucas}(2017)}]{PhysRevB.95.115425}%
  \BibitemOpen
  \bibfield  {author} {\bibinfo {author} {\bibfnamefont {A.}~\bibnamefont
  {Lucas}},\ }\href {\doibase 10.1103/PhysRevB.95.115425} {\bibfield  {journal}
  {\bibinfo  {journal} {Phys. Rev. B}\ }\textbf {\bibinfo {volume} {95}},\
  \bibinfo {pages} {115425} (\bibinfo {year} {2017})}\BibitemShut {NoStop}%
\bibitem [{\citenamefont {Levchenko}\ \emph {et~al.}(2017)\citenamefont
  {Levchenko}, \citenamefont {Xie},\ and\ \citenamefont
  {Andreev}}]{PhysRevB.95.121301}%
  \BibitemOpen
  \bibfield  {author} {\bibinfo {author} {\bibfnamefont {A.}~\bibnamefont
  {Levchenko}}, \bibinfo {author} {\bibfnamefont {H.-Y.}\ \bibnamefont {Xie}},
  \ and\ \bibinfo {author} {\bibfnamefont {A.~V.}\ \bibnamefont {Andreev}},\
  }\href {\doibase 10.1103/PhysRevB.95.121301} {\bibfield  {journal} {\bibinfo
  {journal} {Phys. Rev. B}\ }\textbf {\bibinfo {volume} {95}},\ \bibinfo
  {pages} {121301} (\bibinfo {year} {2017})}\BibitemShut {NoStop}%
\bibitem [{\citenamefont {Bandurin}\ \emph {et~al.}(2016)\citenamefont
  {Bandurin}, \citenamefont {Torre}, \citenamefont {Kumar}, \citenamefont
  {Ben~Shalom}, \citenamefont {Tomadin}, \citenamefont {Principi},
  \citenamefont {Auton}, \citenamefont {Khestanova}, \citenamefont {Novoselov},
  \citenamefont {Grigorieva}, \citenamefont {Ponomarenko}, \citenamefont
  {Geim},\ and\ \citenamefont {Polini}}]{Bandurin1055}%
  \BibitemOpen
  \bibfield  {author} {\bibinfo {author} {\bibfnamefont {D.~A.}\ \bibnamefont
  {Bandurin}}, \bibinfo {author} {\bibfnamefont {I.}~\bibnamefont {Torre}},
  \bibinfo {author} {\bibfnamefont {R.~K.}\ \bibnamefont {Kumar}}, \bibinfo
  {author} {\bibfnamefont {M.}~\bibnamefont {Ben~Shalom}}, \bibinfo {author}
  {\bibfnamefont {A.}~\bibnamefont {Tomadin}}, \bibinfo {author} {\bibfnamefont
  {A.}~\bibnamefont {Principi}}, \bibinfo {author} {\bibfnamefont {G.~H.}\
  \bibnamefont {Auton}}, \bibinfo {author} {\bibfnamefont {E.}~\bibnamefont
  {Khestanova}}, \bibinfo {author} {\bibfnamefont {K.~S.}\ \bibnamefont
  {Novoselov}}, \bibinfo {author} {\bibfnamefont {I.~V.}\ \bibnamefont
  {Grigorieva}}, \bibinfo {author} {\bibfnamefont {L.~A.}\ \bibnamefont
  {Ponomarenko}}, \bibinfo {author} {\bibfnamefont {A.~K.}\ \bibnamefont
  {Geim}}, \ and\ \bibinfo {author} {\bibfnamefont {M.}~\bibnamefont
  {Polini}},\ }\href {\doibase 10.1126/science.aad0201} {\bibfield  {journal}
  {\bibinfo  {journal} {Science}\ }\textbf {\bibinfo {volume} {351}},\ \bibinfo
  {pages} {1055} (\bibinfo {year} {2016})}\BibitemShut {NoStop}%
\bibitem [{\citenamefont {Crossno}\ \emph {et~al.}(2016)\citenamefont
  {Crossno}, \citenamefont {Shi}, \citenamefont {Wang}, \citenamefont {Liu},
  \citenamefont {Harzheim}, \citenamefont {Lucas}, \citenamefont {Sachdev},
  \citenamefont {Kim}, \citenamefont {Taniguchi}, \citenamefont {Watanabe},
  \citenamefont {Ohki},\ and\ \citenamefont {Fong}}]{Crossno1058}%
  \BibitemOpen
  \bibfield  {author} {\bibinfo {author} {\bibfnamefont {J.}~\bibnamefont
  {Crossno}}, \bibinfo {author} {\bibfnamefont {J.~K.}\ \bibnamefont {Shi}},
  \bibinfo {author} {\bibfnamefont {K.}~\bibnamefont {Wang}}, \bibinfo {author}
  {\bibfnamefont {X.}~\bibnamefont {Liu}}, \bibinfo {author} {\bibfnamefont
  {A.}~\bibnamefont {Harzheim}}, \bibinfo {author} {\bibfnamefont
  {A.}~\bibnamefont {Lucas}}, \bibinfo {author} {\bibfnamefont
  {S.}~\bibnamefont {Sachdev}}, \bibinfo {author} {\bibfnamefont
  {P.}~\bibnamefont {Kim}}, \bibinfo {author} {\bibfnamefont {T.}~\bibnamefont
  {Taniguchi}}, \bibinfo {author} {\bibfnamefont {K.}~\bibnamefont {Watanabe}},
  \bibinfo {author} {\bibfnamefont {T.~A.}\ \bibnamefont {Ohki}}, \ and\
  \bibinfo {author} {\bibfnamefont {K.~C.}\ \bibnamefont {Fong}},\ }\href
  {\doibase 10.1126/science.aad0343} {\bibfield  {journal} {\bibinfo  {journal}
  {Science}\ }\textbf {\bibinfo {volume} {351}},\ \bibinfo {pages} {1058}
  (\bibinfo {year} {2016})}\BibitemShut {NoStop}%
\bibitem [{\citenamefont {Moll}\ \emph {et~al.}(2016)\citenamefont {Moll},
  \citenamefont {Kushwaha}, \citenamefont {Nandi}, \citenamefont {Schmidt},\
  and\ \citenamefont {Mackenzie}}]{Moll1061}%
  \BibitemOpen
  \bibfield  {author} {\bibinfo {author} {\bibfnamefont {P.~J.~W.}\
  \bibnamefont {Moll}}, \bibinfo {author} {\bibfnamefont {P.}~\bibnamefont
  {Kushwaha}}, \bibinfo {author} {\bibfnamefont {N.}~\bibnamefont {Nandi}},
  \bibinfo {author} {\bibfnamefont {B.}~\bibnamefont {Schmidt}}, \ and\
  \bibinfo {author} {\bibfnamefont {A.~P.}\ \bibnamefont {Mackenzie}},\ }\href
  {\doibase 10.1126/science.aac8385} {\bibfield  {journal} {\bibinfo  {journal}
  {Science}\ }\textbf {\bibinfo {volume} {351}},\ \bibinfo {pages} {1061}
  (\bibinfo {year} {2016})}\BibitemShut {NoStop}%
\bibitem [{\citenamefont {Narozhny}\ \emph {et~al.}(2017)\citenamefont
  {Narozhny}, \citenamefont {Gornyi}, \citenamefont {Mirlin},\ and\
  \citenamefont {Schmalian}}]{Narozhny2017}%
  \BibitemOpen
  \bibfield  {author} {\bibinfo {author} {\bibfnamefont {B.~N.}\ \bibnamefont
  {Narozhny}}, \bibinfo {author} {\bibfnamefont {I.~V.}\ \bibnamefont
  {Gornyi}}, \bibinfo {author} {\bibfnamefont {A.~D.}\ \bibnamefont {Mirlin}},
  \ and\ \bibinfo {author} {\bibfnamefont {J.}~\bibnamefont {Schmalian}},\
  }\href {\doibase https://doi.org/10.1002/andp.201700043} {\bibfield
  {journal} {\bibinfo  {journal} {Annalen der Physik}\ }\textbf {\bibinfo
  {volume} {529}},\ \bibinfo {pages} {1700043} (\bibinfo {year}
  {2017})}\BibitemShut {NoStop}%
\bibitem [{\citenamefont {Guo}\ \emph {et~al.}(2017)\citenamefont {Guo},
  \citenamefont {Ilseven}, \citenamefont {Falkovich},\ and\ \citenamefont
  {Levitov}}]{Guo3068}%
  \BibitemOpen
  \bibfield  {author} {\bibinfo {author} {\bibfnamefont {H.}~\bibnamefont
  {Guo}}, \bibinfo {author} {\bibfnamefont {E.}~\bibnamefont {Ilseven}},
  \bibinfo {author} {\bibfnamefont {G.}~\bibnamefont {Falkovich}}, \ and\
  \bibinfo {author} {\bibfnamefont {L.~S.}\ \bibnamefont {Levitov}},\ }\href
  {\doibase 10.1073/pnas.1612181114} {\bibfield  {journal} {\bibinfo  {journal}
  {Proceedings of the National Academy of Sciences}\ }\textbf {\bibinfo
  {volume} {114}},\ \bibinfo {pages} {3068} (\bibinfo {year}
  {2017})}\BibitemShut {NoStop}%
\bibitem [{\citenamefont {Krishna~Kumar}\ \emph {et~al.}(2017)\citenamefont
  {Krishna~Kumar}, \citenamefont {Bandurin}, \citenamefont {Pellegrino},
  \citenamefont {Cao}, \citenamefont {Principi}, \citenamefont {Guo},
  \citenamefont {Auton}, \citenamefont {Ben~Shalom}, \citenamefont
  {Ponomarenko}, \citenamefont {Falkovich}, \citenamefont {Watanabe},
  \citenamefont {Taniguchi}, \citenamefont {Grigorieva}, \citenamefont
  {Levitov}, \citenamefont {Polini},\ and\ \citenamefont {Geim}}]{Kumar2017}%
  \BibitemOpen
  \bibfield  {author} {\bibinfo {author} {\bibfnamefont {R.}~\bibnamefont
  {Krishna~Kumar}}, \bibinfo {author} {\bibfnamefont {D.~A.}\ \bibnamefont
  {Bandurin}}, \bibinfo {author} {\bibfnamefont {F.~M.~D.}\ \bibnamefont
  {Pellegrino}}, \bibinfo {author} {\bibfnamefont {Y.}~\bibnamefont {Cao}},
  \bibinfo {author} {\bibfnamefont {A.}~\bibnamefont {Principi}}, \bibinfo
  {author} {\bibfnamefont {H.}~\bibnamefont {Guo}}, \bibinfo {author}
  {\bibfnamefont {G.~H.}\ \bibnamefont {Auton}}, \bibinfo {author}
  {\bibfnamefont {M.}~\bibnamefont {Ben~Shalom}}, \bibinfo {author}
  {\bibfnamefont {L.~A.}\ \bibnamefont {Ponomarenko}}, \bibinfo {author}
  {\bibfnamefont {G.}~\bibnamefont {Falkovich}}, \bibinfo {author}
  {\bibfnamefont {K.}~\bibnamefont {Watanabe}}, \bibinfo {author}
  {\bibfnamefont {T.}~\bibnamefont {Taniguchi}}, \bibinfo {author}
  {\bibfnamefont {I.~V.}\ \bibnamefont {Grigorieva}}, \bibinfo {author}
  {\bibfnamefont {L.~S.}\ \bibnamefont {Levitov}}, \bibinfo {author}
  {\bibfnamefont {M.}~\bibnamefont {Polini}}, \ and\ \bibinfo {author}
  {\bibfnamefont {A.~K.}\ \bibnamefont {Geim}},\ }\href {\doibase
  10.1038/nphys4240} {\bibfield  {journal} {\bibinfo  {journal} {Nature
  Physics}\ }\textbf {\bibinfo {volume} {13}},\ \bibinfo {pages} {1182}
  (\bibinfo {year} {2017})}\BibitemShut {NoStop}%
\bibitem [{\citenamefont {Scaffidi}\ \emph {et~al.}(2017)\citenamefont
  {Scaffidi}, \citenamefont {Nandi}, \citenamefont {Schmidt}, \citenamefont
  {Mackenzie},\ and\ \citenamefont {Moore}}]{Scaffidi2017}%
  \BibitemOpen
  \bibfield  {author} {\bibinfo {author} {\bibfnamefont {T.}~\bibnamefont
  {Scaffidi}}, \bibinfo {author} {\bibfnamefont {N.}~\bibnamefont {Nandi}},
  \bibinfo {author} {\bibfnamefont {B.}~\bibnamefont {Schmidt}}, \bibinfo
  {author} {\bibfnamefont {A.~P.}\ \bibnamefont {Mackenzie}}, \ and\ \bibinfo
  {author} {\bibfnamefont {J.~E.}\ \bibnamefont {Moore}},\ }\href {\doibase
  10.1103/PhysRevLett.118.226601} {\bibfield  {journal} {\bibinfo  {journal}
  {Phys. Rev. Lett.}\ }\textbf {\bibinfo {volume} {118}},\ \bibinfo {pages}
  {226601} (\bibinfo {year} {2017})}\BibitemShut {NoStop}%
\bibitem [{\citenamefont {Lucas}\ and\ \citenamefont
  {Hartnoll}(2018)}]{PhysRevB.97.045105}%
  \BibitemOpen
  \bibfield  {author} {\bibinfo {author} {\bibfnamefont {A.}~\bibnamefont
  {Lucas}}\ and\ \bibinfo {author} {\bibfnamefont {S.~A.}\ \bibnamefont
  {Hartnoll}},\ }\href {\doibase 10.1103/PhysRevB.97.045105} {\bibfield
  {journal} {\bibinfo  {journal} {Phys. Rev. B}\ }\textbf {\bibinfo {volume}
  {97}},\ \bibinfo {pages} {045105} (\bibinfo {year} {2018})}\BibitemShut
  {NoStop}%
\bibitem [{\citenamefont {Ho}\ \emph {et~al.}(2018)\citenamefont {Ho},
  \citenamefont {Yudhistira}, \citenamefont {Chakraborty},\ and\ \citenamefont
  {Adam}}]{PhysRevB.97.121404}%
  \BibitemOpen
  \bibfield  {author} {\bibinfo {author} {\bibfnamefont {D.~Y.~H.}\
  \bibnamefont {Ho}}, \bibinfo {author} {\bibfnamefont {I.}~\bibnamefont
  {Yudhistira}}, \bibinfo {author} {\bibfnamefont {N.}~\bibnamefont
  {Chakraborty}}, \ and\ \bibinfo {author} {\bibfnamefont {S.}~\bibnamefont
  {Adam}},\ }\href {\doibase 10.1103/PhysRevB.97.121404} {\bibfield  {journal}
  {\bibinfo  {journal} {Phys. Rev. B}\ }\textbf {\bibinfo {volume} {97}},\
  \bibinfo {pages} {121404} (\bibinfo {year} {2018})}\BibitemShut {NoStop}%
\bibitem [{\citenamefont {Shytov}\ \emph {et~al.}(2018)\citenamefont {Shytov},
  \citenamefont {Kong}, \citenamefont {Falkovich},\ and\ \citenamefont
  {Levitov}}]{PhysRevLett.121.176805}%
  \BibitemOpen
  \bibfield  {author} {\bibinfo {author} {\bibfnamefont {A.}~\bibnamefont
  {Shytov}}, \bibinfo {author} {\bibfnamefont {J.~F.}\ \bibnamefont {Kong}},
  \bibinfo {author} {\bibfnamefont {G.}~\bibnamefont {Falkovich}}, \ and\
  \bibinfo {author} {\bibfnamefont {L.}~\bibnamefont {Levitov}},\ }\href
  {\doibase 10.1103/PhysRevLett.121.176805} {\bibfield  {journal} {\bibinfo
  {journal} {Phys. Rev. Lett.}\ }\textbf {\bibinfo {volume} {121}},\ \bibinfo
  {pages} {176805} (\bibinfo {year} {2018})}\BibitemShut {NoStop}%
\bibitem [{\citenamefont {Alekseev}\ and\ \citenamefont
  {Semina}(2018)}]{PhysRevB.98.165412}%
  \BibitemOpen
  \bibfield  {author} {\bibinfo {author} {\bibfnamefont {P.~S.}\ \bibnamefont
  {Alekseev}}\ and\ \bibinfo {author} {\bibfnamefont {M.~A.}\ \bibnamefont
  {Semina}},\ }\href {\doibase 10.1103/PhysRevB.98.165412} {\bibfield
  {journal} {\bibinfo  {journal} {Phys. Rev. B}\ }\textbf {\bibinfo {volume}
  {98}},\ \bibinfo {pages} {165412} (\bibinfo {year} {2018})}\BibitemShut
  {NoStop}%
\bibitem [{\citenamefont {Svintsov}(2018)}]{PhysRevB.97.121405}%
  \BibitemOpen
  \bibfield  {author} {\bibinfo {author} {\bibfnamefont {D.}~\bibnamefont
  {Svintsov}},\ }\href {\doibase 10.1103/PhysRevB.97.121405} {\bibfield
  {journal} {\bibinfo  {journal} {Phys. Rev. B}\ }\textbf {\bibinfo {volume}
  {97}},\ \bibinfo {pages} {121405} (\bibinfo {year} {2018})}\BibitemShut
  {NoStop}%
\bibitem [{\citenamefont {Gooth}\ \emph {et~al.}(2018)\citenamefont {Gooth},
  \citenamefont {Menges}, \citenamefont {Kumar}, \citenamefont {S{\"u}ss},
  \citenamefont {Shekhar}, \citenamefont {Sun}, \citenamefont {Drechsler},
  \citenamefont {Zierold}, \citenamefont {Felser},\ and\ \citenamefont
  {Gotsmann}}]{Gooth2018}%
  \BibitemOpen
  \bibfield  {author} {\bibinfo {author} {\bibfnamefont {J.}~\bibnamefont
  {Gooth}}, \bibinfo {author} {\bibfnamefont {F.}~\bibnamefont {Menges}},
  \bibinfo {author} {\bibfnamefont {N.}~\bibnamefont {Kumar}}, \bibinfo
  {author} {\bibfnamefont {V.}~\bibnamefont {S{\"u}ss}}, \bibinfo {author}
  {\bibfnamefont {C.}~\bibnamefont {Shekhar}}, \bibinfo {author} {\bibfnamefont
  {Y.}~\bibnamefont {Sun}}, \bibinfo {author} {\bibfnamefont {U.}~\bibnamefont
  {Drechsler}}, \bibinfo {author} {\bibfnamefont {R.}~\bibnamefont {Zierold}},
  \bibinfo {author} {\bibfnamefont {C.}~\bibnamefont {Felser}}, \ and\ \bibinfo
  {author} {\bibfnamefont {B.}~\bibnamefont {Gotsmann}},\ }\href {\doibase
  10.1038/s41467-018-06688-y} {\bibfield  {journal} {\bibinfo  {journal}
  {Nature Communications}\ }\textbf {\bibinfo {volume} {9}},\ \bibinfo {pages}
  {4093} (\bibinfo {year} {2018})}\BibitemShut {NoStop}%
\bibitem [{\citenamefont {Braem}\ \emph {et~al.}(2018)\citenamefont {Braem},
  \citenamefont {Pellegrino}, \citenamefont {Principi}, \citenamefont
  {R\"o\"osli}, \citenamefont {Gold}, \citenamefont {Hennel}, \citenamefont
  {Koski}, \citenamefont {Berl}, \citenamefont {Dietsche}, \citenamefont
  {Wegscheider}, \citenamefont {Polini}, \citenamefont {Ihn},\ and\
  \citenamefont {Ensslin}}]{Braem2018}%
  \BibitemOpen
  \bibfield  {author} {\bibinfo {author} {\bibfnamefont {B.~A.}\ \bibnamefont
  {Braem}}, \bibinfo {author} {\bibfnamefont {F.~M.~D.}\ \bibnamefont
  {Pellegrino}}, \bibinfo {author} {\bibfnamefont {A.}~\bibnamefont
  {Principi}}, \bibinfo {author} {\bibfnamefont {M.}~\bibnamefont
  {R\"o\"osli}}, \bibinfo {author} {\bibfnamefont {C.}~\bibnamefont {Gold}},
  \bibinfo {author} {\bibfnamefont {S.}~\bibnamefont {Hennel}}, \bibinfo
  {author} {\bibfnamefont {J.~V.}\ \bibnamefont {Koski}}, \bibinfo {author}
  {\bibfnamefont {M.}~\bibnamefont {Berl}}, \bibinfo {author} {\bibfnamefont
  {W.}~\bibnamefont {Dietsche}}, \bibinfo {author} {\bibfnamefont
  {W.}~\bibnamefont {Wegscheider}}, \bibinfo {author} {\bibfnamefont
  {M.}~\bibnamefont {Polini}}, \bibinfo {author} {\bibfnamefont
  {T.}~\bibnamefont {Ihn}}, \ and\ \bibinfo {author} {\bibfnamefont
  {K.}~\bibnamefont {Ensslin}},\ }\href {\doibase 10.1103/PhysRevB.98.241304}
  {\bibfield  {journal} {\bibinfo  {journal} {Phys. Rev. B}\ }\textbf {\bibinfo
  {volume} {98}},\ \bibinfo {pages} {241304} (\bibinfo {year}
  {2018})}\BibitemShut {NoStop}%
\bibitem [{\citenamefont {{Berdyugin}}\ \emph {et~al.}(2019)\citenamefont
  {{Berdyugin}}, \citenamefont {{Xu}}, \citenamefont {{Pellegrino}},
  \citenamefont {{Krishna Kumar}}, \citenamefont {{Principi}}, \citenamefont
  {{Torre}}, \citenamefont {{Ben Shalom}}, \citenamefont {{Taniguchi}},
  \citenamefont {{Watanabe}}, \citenamefont {{Grigorieva}}, \citenamefont
  {{Polini}}, \citenamefont {{Geim}},\ and\ \citenamefont
  {{Bandurin}}}]{Berdyugin2019}%
  \BibitemOpen
  \bibfield  {author} {\bibinfo {author} {\bibfnamefont {A.~I.}\ \bibnamefont
  {{Berdyugin}}}, \bibinfo {author} {\bibfnamefont {S.~G.}\ \bibnamefont
  {{Xu}}}, \bibinfo {author} {\bibfnamefont {F.~M.~D.}\ \bibnamefont
  {{Pellegrino}}}, \bibinfo {author} {\bibfnamefont {R.}~\bibnamefont {{Krishna
  Kumar}}}, \bibinfo {author} {\bibfnamefont {A.}~\bibnamefont {{Principi}}},
  \bibinfo {author} {\bibfnamefont {I.}~\bibnamefont {{Torre}}}, \bibinfo
  {author} {\bibfnamefont {M.}~\bibnamefont {{Ben Shalom}}}, \bibinfo {author}
  {\bibfnamefont {T.}~\bibnamefont {{Taniguchi}}}, \bibinfo {author}
  {\bibfnamefont {K.}~\bibnamefont {{Watanabe}}}, \bibinfo {author}
  {\bibfnamefont {I.~V.}\ \bibnamefont {{Grigorieva}}}, \bibinfo {author}
  {\bibfnamefont {M.}~\bibnamefont {{Polini}}}, \bibinfo {author}
  {\bibfnamefont {A.~K.}\ \bibnamefont {{Geim}}}, \ and\ \bibinfo {author}
  {\bibfnamefont {D.~A.}\ \bibnamefont {{Bandurin}}},\ }\href {\doibase
  10.1126/science.aau0685} {\bibfield  {journal} {\bibinfo  {journal}
  {Science}\ }\textbf {\bibinfo {volume} {364}},\ \bibinfo {pages} {162}
  (\bibinfo {year} {2019})},\ \Eprint {http://arxiv.org/abs/1806.01606}
  {arXiv:1806.01606 [cond-mat.mes-hall]} \BibitemShut {NoStop}%
\bibitem [{\citenamefont {Sulpizio}\ \emph {et~al.}(2019)\citenamefont
  {Sulpizio}, \citenamefont {Ella}, \citenamefont {Rozen}, \citenamefont
  {Birkbeck}, \citenamefont {Perello}, \citenamefont {Dutta}, \citenamefont
  {Ben-Shalom}, \citenamefont {Taniguchi}, \citenamefont {Watanabe},
  \citenamefont {Holder}, \citenamefont {Queiroz}, \citenamefont {Principi},
  \citenamefont {Stern}, \citenamefont {Scaffidi}, \citenamefont {Geim},\ and\
  \citenamefont {Ilani}}]{Sulpizio2019}%
  \BibitemOpen
  \bibfield  {author} {\bibinfo {author} {\bibfnamefont {J.~A.}\ \bibnamefont
  {Sulpizio}}, \bibinfo {author} {\bibfnamefont {L.}~\bibnamefont {Ella}},
  \bibinfo {author} {\bibfnamefont {A.}~\bibnamefont {Rozen}}, \bibinfo
  {author} {\bibfnamefont {J.}~\bibnamefont {Birkbeck}}, \bibinfo {author}
  {\bibfnamefont {D.~J.}\ \bibnamefont {Perello}}, \bibinfo {author}
  {\bibfnamefont {D.}~\bibnamefont {Dutta}}, \bibinfo {author} {\bibfnamefont
  {M.}~\bibnamefont {Ben-Shalom}}, \bibinfo {author} {\bibfnamefont
  {T.}~\bibnamefont {Taniguchi}}, \bibinfo {author} {\bibfnamefont
  {K.}~\bibnamefont {Watanabe}}, \bibinfo {author} {\bibfnamefont
  {T.}~\bibnamefont {Holder}}, \bibinfo {author} {\bibfnamefont
  {R.}~\bibnamefont {Queiroz}}, \bibinfo {author} {\bibfnamefont
  {A.}~\bibnamefont {Principi}}, \bibinfo {author} {\bibfnamefont
  {A.}~\bibnamefont {Stern}}, \bibinfo {author} {\bibfnamefont
  {T.}~\bibnamefont {Scaffidi}}, \bibinfo {author} {\bibfnamefont {A.~K.}\
  \bibnamefont {Geim}}, \ and\ \bibinfo {author} {\bibfnamefont
  {S.}~\bibnamefont {Ilani}},\ }\href {\doibase 10.1038/s41586-019-1788-9}
  {\bibfield  {journal} {\bibinfo  {journal} {Nature}\ }\textbf {\bibinfo
  {volume} {576}},\ \bibinfo {pages} {75} (\bibinfo {year} {2019})}\BibitemShut
  {NoStop}%
\bibitem [{\citenamefont {Narozhny}(2019)}]{NAROZHNY2019167979}%
  \BibitemOpen
  \bibfield  {author} {\bibinfo {author} {\bibfnamefont {B.~N.}\ \bibnamefont
  {Narozhny}},\ }\href {\doibase https://doi.org/10.1016/j.aop.2019.167979}
  {\bibfield  {journal} {\bibinfo  {journal} {Annals of Physics}\ }\textbf
  {\bibinfo {volume} {411}},\ \bibinfo {pages} {167979} (\bibinfo {year}
  {2019})}\BibitemShut {NoStop}%
\bibitem [{\citenamefont {Shavit}\ \emph {et~al.}(2019)\citenamefont {Shavit},
  \citenamefont {Shytov},\ and\ \citenamefont
  {Falkovich}}]{PhysRevLett.123.026801}%
  \BibitemOpen
  \bibfield  {author} {\bibinfo {author} {\bibfnamefont {M.}~\bibnamefont
  {Shavit}}, \bibinfo {author} {\bibfnamefont {A.}~\bibnamefont {Shytov}}, \
  and\ \bibinfo {author} {\bibfnamefont {G.}~\bibnamefont {Falkovich}},\ }\href
  {\doibase 10.1103/PhysRevLett.123.026801} {\bibfield  {journal} {\bibinfo
  {journal} {Phys. Rev. Lett.}\ }\textbf {\bibinfo {volume} {123}},\ \bibinfo
  {pages} {026801} (\bibinfo {year} {2019})}\BibitemShut {NoStop}%
\bibitem [{\citenamefont {Ku}\ \emph {et~al.}(2020)\citenamefont {Ku},
  \citenamefont {Zhou}, \citenamefont {Li}, \citenamefont {Shin}, \citenamefont
  {Shi}, \citenamefont {Burch}, \citenamefont {Anderson}, \citenamefont
  {Pierce}, \citenamefont {Xie}, \citenamefont {Hamo}, \citenamefont {Vool},
  \citenamefont {Zhang}, \citenamefont {Casola}, \citenamefont {Taniguchi},
  \citenamefont {Watanabe}, \citenamefont {Fogler}, \citenamefont {Kim},
  \citenamefont {Yacoby},\ and\ \citenamefont {Walsworth}}]{Ku2020}%
  \BibitemOpen
  \bibfield  {author} {\bibinfo {author} {\bibfnamefont {M.~J.~H.}\
  \bibnamefont {Ku}}, \bibinfo {author} {\bibfnamefont {T.~X.}\ \bibnamefont
  {Zhou}}, \bibinfo {author} {\bibfnamefont {Q.}~\bibnamefont {Li}}, \bibinfo
  {author} {\bibfnamefont {Y.~J.}\ \bibnamefont {Shin}}, \bibinfo {author}
  {\bibfnamefont {J.~K.}\ \bibnamefont {Shi}}, \bibinfo {author} {\bibfnamefont
  {C.}~\bibnamefont {Burch}}, \bibinfo {author} {\bibfnamefont {L.~E.}\
  \bibnamefont {Anderson}}, \bibinfo {author} {\bibfnamefont {A.~T.}\
  \bibnamefont {Pierce}}, \bibinfo {author} {\bibfnamefont {Y.}~\bibnamefont
  {Xie}}, \bibinfo {author} {\bibfnamefont {A.}~\bibnamefont {Hamo}}, \bibinfo
  {author} {\bibfnamefont {U.}~\bibnamefont {Vool}}, \bibinfo {author}
  {\bibfnamefont {H.}~\bibnamefont {Zhang}}, \bibinfo {author} {\bibfnamefont
  {F.}~\bibnamefont {Casola}}, \bibinfo {author} {\bibfnamefont
  {T.}~\bibnamefont {Taniguchi}}, \bibinfo {author} {\bibfnamefont
  {K.}~\bibnamefont {Watanabe}}, \bibinfo {author} {\bibfnamefont {M.~M.}\
  \bibnamefont {Fogler}}, \bibinfo {author} {\bibfnamefont {P.}~\bibnamefont
  {Kim}}, \bibinfo {author} {\bibfnamefont {A.}~\bibnamefont {Yacoby}}, \ and\
  \bibinfo {author} {\bibfnamefont {R.~L.}\ \bibnamefont {Walsworth}},\ }\href
  {\doibase 10.1038/s41586-020-2507-2} {\bibfield  {journal} {\bibinfo
  {journal} {Nature}\ }\textbf {\bibinfo {volume} {583}},\ \bibinfo {pages}
  {537} (\bibinfo {year} {2020})}\BibitemShut {NoStop}%
\bibitem [{\citenamefont {Levchenko}\ and\ \citenamefont
  {Schmalian}(2020)}]{LEVCHENKO2020168218}%
  \BibitemOpen
  \bibfield  {author} {\bibinfo {author} {\bibfnamefont {A.}~\bibnamefont
  {Levchenko}}\ and\ \bibinfo {author} {\bibfnamefont {J.}~\bibnamefont
  {Schmalian}},\ }\href {\doibase https://doi.org/10.1016/j.aop.2020.168218}
  {\bibfield  {journal} {\bibinfo  {journal} {Annals of Physics}\ }\textbf
  {\bibinfo {volume} {419}},\ \bibinfo {pages} {168218} (\bibinfo {year}
  {2020})}\BibitemShut {NoStop}%
\bibitem [{\citenamefont {Holder}\ \emph {et~al.}(2019)\citenamefont {Holder},
  \citenamefont {Queiroz}, \citenamefont {Scaffidi}, \citenamefont
  {Silberstein}, \citenamefont {Rozen}, \citenamefont {Sulpizio}, \citenamefont
  {Ella}, \citenamefont {Ilani},\ and\ \citenamefont
  {Stern}}]{PhysRevB.100.245305}%
  \BibitemOpen
  \bibfield  {author} {\bibinfo {author} {\bibfnamefont {T.}~\bibnamefont
  {Holder}}, \bibinfo {author} {\bibfnamefont {R.}~\bibnamefont {Queiroz}},
  \bibinfo {author} {\bibfnamefont {T.}~\bibnamefont {Scaffidi}}, \bibinfo
  {author} {\bibfnamefont {N.}~\bibnamefont {Silberstein}}, \bibinfo {author}
  {\bibfnamefont {A.}~\bibnamefont {Rozen}}, \bibinfo {author} {\bibfnamefont
  {J.~A.}\ \bibnamefont {Sulpizio}}, \bibinfo {author} {\bibfnamefont
  {L.}~\bibnamefont {Ella}}, \bibinfo {author} {\bibfnamefont {S.}~\bibnamefont
  {Ilani}}, \ and\ \bibinfo {author} {\bibfnamefont {A.}~\bibnamefont
  {Stern}},\ }\href {\doibase 10.1103/PhysRevB.100.245305} {\bibfield
  {journal} {\bibinfo  {journal} {Phys. Rev. B}\ }\textbf {\bibinfo {volume}
  {100}},\ \bibinfo {pages} {245305} (\bibinfo {year} {2019})}\BibitemShut
  {NoStop}%
\bibitem [{\citenamefont {{Jenkins}}\ \emph {et~al.}(2020)\citenamefont
  {{Jenkins}}, \citenamefont {{Baumann}}, \citenamefont {{Zhou}}, \citenamefont
  {{Meynell}}, \citenamefont {{Yang}}, \citenamefont {{Watanabe}},
  \citenamefont {{Taniguchi}}, \citenamefont {{Lucas}}, \citenamefont
  {{Young}},\ and\ \citenamefont {{Bleszynski Jayich}}}]{Jenkins2020}%
  \BibitemOpen
  \bibfield  {author} {\bibinfo {author} {\bibfnamefont {A.}~\bibnamefont
  {{Jenkins}}}, \bibinfo {author} {\bibfnamefont {S.}~\bibnamefont
  {{Baumann}}}, \bibinfo {author} {\bibfnamefont {H.}~\bibnamefont {{Zhou}}},
  \bibinfo {author} {\bibfnamefont {S.~A.}\ \bibnamefont {{Meynell}}}, \bibinfo
  {author} {\bibfnamefont {D.}~\bibnamefont {{Yang}}}, \bibinfo {author}
  {\bibfnamefont {K.}~\bibnamefont {{Watanabe}}}, \bibinfo {author}
  {\bibfnamefont {T.}~\bibnamefont {{Taniguchi}}}, \bibinfo {author}
  {\bibfnamefont {A.}~\bibnamefont {{Lucas}}}, \bibinfo {author} {\bibfnamefont
  {A.~F.}\ \bibnamefont {{Young}}}, \ and\ \bibinfo {author} {\bibfnamefont
  {A.~C.}\ \bibnamefont {{Bleszynski Jayich}}},\ }\href@noop {} {\bibfield
  {journal} {\bibinfo  {journal} {arXiv e-prints}\ ,\ \bibinfo {eid}
  {arXiv:2002.05065}} (\bibinfo {year} {2020})},\ \Eprint
  {http://arxiv.org/abs/2002.05065} {arXiv:2002.05065 [cond-mat.mes-hall]}
  \BibitemShut {NoStop}%
\bibitem [{\citenamefont {Keser}\ \emph {et~al.}(2021)\citenamefont {Keser},
  \citenamefont {Wang}, \citenamefont {Klochan}, \citenamefont {Ho},
  \citenamefont {Tkachenko}, \citenamefont {Tkachenko}, \citenamefont {Culcer},
  \citenamefont {Adam}, \citenamefont {Farrer}, \citenamefont {Ritchie},
  \citenamefont {Sushkov},\ and\ \citenamefont {Hamilton}}]{Keser2021}%
  \BibitemOpen
  \bibfield  {author} {\bibinfo {author} {\bibfnamefont {A.~C.}\ \bibnamefont
  {Keser}}, \bibinfo {author} {\bibfnamefont {D.~Q.}\ \bibnamefont {Wang}},
  \bibinfo {author} {\bibfnamefont {O.}~\bibnamefont {Klochan}}, \bibinfo
  {author} {\bibfnamefont {D.~Y.~H.}\ \bibnamefont {Ho}}, \bibinfo {author}
  {\bibfnamefont {O.~A.}\ \bibnamefont {Tkachenko}}, \bibinfo {author}
  {\bibfnamefont {V.~A.}\ \bibnamefont {Tkachenko}}, \bibinfo {author}
  {\bibfnamefont {D.}~\bibnamefont {Culcer}}, \bibinfo {author} {\bibfnamefont
  {S.}~\bibnamefont {Adam}}, \bibinfo {author} {\bibfnamefont {I.}~\bibnamefont
  {Farrer}}, \bibinfo {author} {\bibfnamefont {D.~A.}\ \bibnamefont {Ritchie}},
  \bibinfo {author} {\bibfnamefont {O.~P.}\ \bibnamefont {Sushkov}}, \ and\
  \bibinfo {author} {\bibfnamefont {A.~R.}\ \bibnamefont {Hamilton}},\ }\href
  {\doibase 10.1103/PhysRevX.11.031030} {\bibfield  {journal} {\bibinfo
  {journal} {Phys. Rev. X}\ }\textbf {\bibinfo {volume} {11}},\ \bibinfo
  {pages} {031030} (\bibinfo {year} {2021})}\BibitemShut {NoStop}%
\bibitem [{\citenamefont {Gupta}\ \emph {et~al.}(2021)\citenamefont {Gupta},
  \citenamefont {Heremans}, \citenamefont {Kataria}, \citenamefont {Chandra},
  \citenamefont {Fallahi}, \citenamefont {Gardner},\ and\ \citenamefont
  {Manfra}}]{Gupta2021}%
  \BibitemOpen
  \bibfield  {author} {\bibinfo {author} {\bibfnamefont {A.}~\bibnamefont
  {Gupta}}, \bibinfo {author} {\bibfnamefont {J.~J.}\ \bibnamefont {Heremans}},
  \bibinfo {author} {\bibfnamefont {G.}~\bibnamefont {Kataria}}, \bibinfo
  {author} {\bibfnamefont {M.}~\bibnamefont {Chandra}}, \bibinfo {author}
  {\bibfnamefont {S.}~\bibnamefont {Fallahi}}, \bibinfo {author} {\bibfnamefont
  {G.~C.}\ \bibnamefont {Gardner}}, \ and\ \bibinfo {author} {\bibfnamefont
  {M.~J.}\ \bibnamefont {Manfra}},\ }\href {\doibase
  10.1103/PhysRevLett.126.076803} {\bibfield  {journal} {\bibinfo  {journal}
  {Phys. Rev. Lett.}\ }\textbf {\bibinfo {volume} {126}},\ \bibinfo {pages}
  {076803} (\bibinfo {year} {2021})}\BibitemShut {NoStop}%
\bibitem [{\citenamefont {{Krebs}}\ \emph {et~al.}(2021)\citenamefont
  {{Krebs}}, \citenamefont {{Behn}}, \citenamefont {{Li}}, \citenamefont
  {{Smith}}, \citenamefont {{Watanabe}}, \citenamefont {{Taniguchi}},
  \citenamefont {{Levchenko}},\ and\ \citenamefont {{Brar}}}]{Krebs2021}%
  \BibitemOpen
  \bibfield  {author} {\bibinfo {author} {\bibfnamefont {Z.~J.}\ \bibnamefont
  {{Krebs}}}, \bibinfo {author} {\bibfnamefont {W.~A.}\ \bibnamefont {{Behn}}},
  \bibinfo {author} {\bibfnamefont {S.}~\bibnamefont {{Li}}}, \bibinfo {author}
  {\bibfnamefont {K.~J.}\ \bibnamefont {{Smith}}}, \bibinfo {author}
  {\bibfnamefont {K.}~\bibnamefont {{Watanabe}}}, \bibinfo {author}
  {\bibfnamefont {T.}~\bibnamefont {{Taniguchi}}}, \bibinfo {author}
  {\bibfnamefont {A.}~\bibnamefont {{Levchenko}}}, \ and\ \bibinfo {author}
  {\bibfnamefont {V.~W.}\ \bibnamefont {{Brar}}},\ }\href@noop {} {\bibfield
  {journal} {\bibinfo  {journal} {arXiv e-prints}\ ,\ \bibinfo {eid}
  {arXiv:2106.07212}} (\bibinfo {year} {2021})},\ \Eprint
  {http://arxiv.org/abs/2106.07212} {arXiv:2106.07212 [cond-mat.mes-hall]}
  \BibitemShut {NoStop}%
\bibitem [{\citenamefont {{Hong}}\ \emph {et~al.}(2020)\citenamefont {{Hong}},
  \citenamefont {{Davydova}}, \citenamefont {{Ledwith}},\ and\ \citenamefont
  {{Levitov}}}]{Hong2020}%
  \BibitemOpen
  \bibfield  {author} {\bibinfo {author} {\bibfnamefont {Q.}~\bibnamefont
  {{Hong}}}, \bibinfo {author} {\bibfnamefont {M.}~\bibnamefont {{Davydova}}},
  \bibinfo {author} {\bibfnamefont {P.~J.}\ \bibnamefont {{Ledwith}}}, \ and\
  \bibinfo {author} {\bibfnamefont {L.}~\bibnamefont {{Levitov}}},\ }\href@noop
  {} {\bibfield  {journal} {\bibinfo  {journal} {arXiv e-prints}\ ,\ \bibinfo
  {eid} {arXiv:2012.03840}} (\bibinfo {year} {2020})},\ \Eprint
  {http://arxiv.org/abs/2012.03840} {arXiv:2012.03840 [cond-mat.mes-hall]}
  \BibitemShut {NoStop}%
\bibitem [{\citenamefont {Nagaev}\ and\ \citenamefont
  {Ayvazyan}(2008)}]{Nagaev2008}%
  \BibitemOpen
  \bibfield  {author} {\bibinfo {author} {\bibfnamefont {K.~E.}\ \bibnamefont
  {Nagaev}}\ and\ \bibinfo {author} {\bibfnamefont {O.~S.}\ \bibnamefont
  {Ayvazyan}},\ }\href {\doibase 10.1103/PhysRevLett.101.216807} {\bibfield
  {journal} {\bibinfo  {journal} {Phys. Rev. Lett.}\ }\textbf {\bibinfo
  {volume} {101}},\ \bibinfo {pages} {216807} (\bibinfo {year}
  {2008})}\BibitemShut {NoStop}%
\bibitem [{\citenamefont {Nagaev}\ and\ \citenamefont
  {Kostyuchenko}(2010)}]{Nagaev2010}%
  \BibitemOpen
  \bibfield  {author} {\bibinfo {author} {\bibfnamefont {K.~E.}\ \bibnamefont
  {Nagaev}}\ and\ \bibinfo {author} {\bibfnamefont {T.~V.}\ \bibnamefont
  {Kostyuchenko}},\ }\href {\doibase 10.1103/PhysRevB.81.125316} {\bibfield
  {journal} {\bibinfo  {journal} {Phys. Rev. B}\ }\textbf {\bibinfo {volume}
  {81}},\ \bibinfo {pages} {125316} (\bibinfo {year} {2010})}\BibitemShut
  {NoStop}%
\bibitem [{\citenamefont {Melnikov}\ \emph {et~al.}(2012)\citenamefont
  {Melnikov}, \citenamefont {Kotthaus}, \citenamefont {Pellegrini},
  \citenamefont {Sorba}, \citenamefont {Biasiol},\ and\ \citenamefont
  {Khrapai}}]{Melnikov2012}%
  \BibitemOpen
  \bibfield  {author} {\bibinfo {author} {\bibfnamefont {M.~Y.}\ \bibnamefont
  {Melnikov}}, \bibinfo {author} {\bibfnamefont {J.~P.}\ \bibnamefont
  {Kotthaus}}, \bibinfo {author} {\bibfnamefont {V.}~\bibnamefont
  {Pellegrini}}, \bibinfo {author} {\bibfnamefont {L.}~\bibnamefont {Sorba}},
  \bibinfo {author} {\bibfnamefont {G.}~\bibnamefont {Biasiol}}, \ and\
  \bibinfo {author} {\bibfnamefont {V.~S.}\ \bibnamefont {Khrapai}},\ }\href
  {\doibase 10.1103/PhysRevB.86.075425} {\bibfield  {journal} {\bibinfo
  {journal} {Phys. Rev. B}\ }\textbf {\bibinfo {volume} {86}},\ \bibinfo
  {pages} {075425} (\bibinfo {year} {2012})}\BibitemShut {NoStop}%
\bibitem [{\citenamefont {Callaway}(1959)}]{Callaway}%
  \BibitemOpen
  \bibfield  {author} {\bibinfo {author} {\bibfnamefont {J.}~\bibnamefont
  {Callaway}},\ }\href {\doibase 10.1103/PhysRev.113.1046} {\bibfield
  {journal} {\bibinfo  {journal} {Phys. Rev.}\ }\textbf {\bibinfo {volume}
  {113}},\ \bibinfo {pages} {1046} (\bibinfo {year} {1959})}\BibitemShut
  {NoStop}%
\bibitem [{\citenamefont {Kumar}\ \emph {et~al.}(2022)\citenamefont {Kumar},
  \citenamefont {Birkbeck}, \citenamefont {Sulpizio}, \citenamefont {Perello},
  \citenamefont {Taniguchi}, \citenamefont {Watanabe}, \citenamefont {Reuven},
  \citenamefont {Scaffidi}, \citenamefont {Stern}, \citenamefont {Geim},\ and\
  \citenamefont {Ilani}}]{companion}%
  \BibitemOpen
  \bibfield  {author} {\bibinfo {author} {\bibfnamefont {C.}~\bibnamefont
  {Kumar}}, \bibinfo {author} {\bibfnamefont {J.}~\bibnamefont {Birkbeck}},
  \bibinfo {author} {\bibfnamefont {J.~A.}\ \bibnamefont {Sulpizio}}, \bibinfo
  {author} {\bibfnamefont {D.}~\bibnamefont {Perello}}, \bibinfo {author}
  {\bibfnamefont {T.}~\bibnamefont {Taniguchi}}, \bibinfo {author}
  {\bibfnamefont {K.}~\bibnamefont {Watanabe}}, \bibinfo {author}
  {\bibfnamefont {O.}~\bibnamefont {Reuven}}, \bibinfo {author} {\bibfnamefont
  {T.}~\bibnamefont {Scaffidi}}, \bibinfo {author} {\bibfnamefont
  {A.}~\bibnamefont {Stern}}, \bibinfo {author} {\bibfnamefont {A.~K.}\
  \bibnamefont {Geim}}, \ and\ \bibinfo {author} {\bibfnamefont
  {S.}~\bibnamefont {Ilani}},\ }\href {\doibase 10.1038/s41586-022-05002-7}
  {\bibfield  {journal} {\bibinfo  {journal} {Nature}\ }\textbf {\bibinfo
  {volume} {609}},\ \bibinfo {pages} {276} (\bibinfo {year}
  {2022})}\BibitemShut {NoStop}%
\end{thebibliography}%
